\documentclass[acmsmall,screen,nonacm]{acmart}
\usepackage{amsmath}
\usepackage{booktabs,tabularx,array}
\usepackage{graphicx}
\usepackage{xurl}
\usepackage{xcolor}
\usepackage{xspace}
\usepackage{tikz}
\usetikzlibrary{arrows.meta,positioning,fit,decorations.pathreplacing}
\usepackage{pgfplots}
\usepackage{algorithm}
\usepackage[noend]{algpseudocode}
\pgfplotsset{compat=1.18}
\newcommand{\pending}[1]{\textcolor{red}{[PENDING\if\relax\detokenize{#1}\relax\else: #1\fi]}}

\newcommand{\tool}{AdaT$^2$\xspace}
\newcolumntype{Y}{>{\raggedright\arraybackslash}X}

\begin{document}
\title[AdaT$^2$: Adaptive Test Transformations for Boundary Testing]{\texorpdfstring{\tool}{AdaT²}: Adaptive Test Transformations for Black-Box Boundary Testing of Conversational Agents}

\author{Liting Lin}
\affiliation{%
  \institution{Lero, the Research Ireland Centre for Software, University of Limerick}
  \city{Limerick}
  \country{Ireland}}
\email{Liting.Lin@ul.ie}

\author{Boxi Yu}
\affiliation{%
  \institution{Lero, the Research Ireland Centre for Software, University of Limerick}
  \city{Limerick}
  \country{Ireland}}
\email{boxi.yu@lero.ie}

\author{Qinghua Xu}
\affiliation{%
  \institution{Lero, the Research Ireland Centre for Software, University of Limerick}
  \city{Limerick}
  \country{Ireland}}
\email{Qinghua.Xu@ul.ie}

\author{Yuzhong Zhang}
\affiliation{%
  \institution{The Chinese University of Hong Kong, Shenzhen}
  \city{Shenzhen}
  \country{China}}
\email{123090848@link.cuhk.edu.cn}

\author{Lionel Briand}
\affiliation{%
  \institution{Lero, the Research Ireland Centre for Software, University of Limerick}
  \city{Limerick}
  \country{Ireland}}
\affiliation{%
  \institution{University of Ottawa}
  \city{Ottawa}
  \country{Canada}}
\email{lionel.briand@lero.ie}

\author{Emir Mu\~{n}oz}
\affiliation{%
  \institution{Genesys}
  \city{Galway}
  \country{Ireland}}
\email{emir.munoz@gmail.com}

\renewcommand{\shortauthors}{Lin et al.}

\authorsaddresses{Authors' Contact Information: Liting Lin, \href{mailto:Liting.Lin@ul.ie}{Liting.Lin@ul.ie}; Boxi Yu, \href{mailto:boxi.yu@lero.ie}{boxi.yu@lero.ie}; Qinghua Xu, \href{mailto:Qinghua.Xu@ul.ie}{Qinghua.Xu@ul.ie}; Yuzhong Zhang, \href{mailto:123090848@link.cuhk.edu.cn}{123090848@link.cuhk.edu.cn}; Lionel Briand, \href{mailto:lionel.briand@lero.ie}{lionel.briand@lero.ie}; Emir Mu\~{n}oz, \href{mailto:emir.munoz@gmail.com}{emir.munoz@gmail.com}}

\begin{abstract}
Conversational agents based on large language models (LLMs) must comply with policies.
Each condition in a policy draws a boundary between user requests, and the agent must behave differently on the two sides of the boundary.
We present \tool, which extracts statements from the agent's replies in exploratory conversations with an LLM acting as the user, and uses the statements to guide boundary test generation.
Each statement describes one condition and the behavior expected when the condition holds.
Besides plain tests guided by single statements, \tool writes transformed tests guided by pairs of a statement and a test transformation instruction, such as ``omit one required input''.
Each statement describes one condition and the behavior expected when the condition holds.
Besides plain tests for single statements, \tool writes transformed tests guided by pairs of a statement and a test transformation instruction.
The instruction of a pair can move a test to the other side of the statement's boundary or to another boundary.
The statements and instructions form far more pairs than a run can try, and many of these pairs are not applicable.
Adaptive pair selection therefore chooses the statement of each pair by novelty and the instruction with the bandit algorithm Bayes-UCB.
Bayes-UCB learns which instructions to prefer from the outcomes of previously tried pairs: whether they yielded a test and whether the agent passed that test according to an LLM judge.
Our benchmark counts the two sides of each boundary separately and distinguishes boundaries explicitly defined by the agent's prompt, tool code, or knowledge base from boundaries that the LLM driving the agent infers from its domain knowledge.
On four domains of $\tau^3$-bench, 62.7\% to 83.3\% of \tool's tests are valid boundary tests whose expected behavior is explicitly defined.
This proportion is higher in every domain than in previous AgentEval's test suites (47.7\% to 68.1\%).
The transformed tests contribute 13 to 46 explicitly defined boundaries that the plain tests miss.
As regression tests, \tool's test suites detect all eight seeded policy faults in the airline domain and four of eight in the retail domain, and AgentEval's test suites, which have fewer than a third as many tests, detect five and two, respectively.

\end{abstract}

\maketitle

% ======================================================================
% 1 Introduction
% ======================================================================
\section{Introduction}
\label{sec:intro}

Conversational agents based on large language models (LLMs) help users with tasks such as changing a booking or returning an item. They carry out these tasks by making tool calls under prompt instructions~\cite{yao2024taubench,barres2025tau2bench,arcadinho2024almita,yehudai2026survey}.
Throughout a multi-turn conversation, such an agent must comply with its \emph{policy}.
The policy consists of conditions and procedures---such as eligibility limits, required confirmations, and identity checks---that are stated in the agent's prompt, in the knowledge base the agent consults, or in the code of its tools~\cite{yao2024taubench,nakash2025redteaming,balaji2026beyondivr}.
However, evaluations on the benchmarks $\tau$-bench~\cite{yao2024taubench}, CRMArena-Pro~\cite{huang2026crmarenapro}, and SOPBench~\cite{li2025sopbench} show that agents often fail to comply with their policies: they behave inconsistently on repeated trials, lose accuracy over multiple turns, and fail many tests of whether they follow prescribed procedures.
Such failures may be costly, because an action taken through a tool, such as a refund or a cancellation, may not be reversible.

To detect these failures, this paper proposes a method that lets a black-box tester generate boundary tests from the agent's replies, using only the observed conversation.
Boundary tests come from program testing, where domain testing and boundary value analysis place tests on and near the boundaries of a program's input domain to find errors there~\cite{clarke1982domain,hierons2006bva}.
A conversational agent's input consists of user requests, so we define an agent's \emph{boundary} as a division of these requests into two groups, its \emph{sides}, that the agent must handle differently.
The conditions of the agent's policy define such boundaries: requests that satisfy a condition lie on one side, and requests that violate it lie on the other.
For example, the banking agent policy in $\tau^3$-bench~\cite{tau3bench}, a benchmark for conversational agents, allows the agent to change the user's account only after the user correctly provides two of four details: date of birth, email, phone number, and address.
If a user asks the agent to raise the daily ATM limit on a debit card and provides the date of birth and phone number, the request belongs to the side where the agent may raise the limit.
If the user provides only the date of birth, the request belongs to the side where the agent must not raise the limit until the user provides a second detail; an agent that raises the limit in this case fails to comply with the condition (Figure~\ref{fig:motivation}, right).
In a \emph{boundary test}, the tester sends the agent a request from one side of a boundary and checks whether the agent exhibits the behavior expected on that side; thus, boundary tests directly assess whether the agent complies with the specified conditions.

\usetikzlibrary{shadows}

\begin{figure}[t]
\centering
% Shared palette (Okabe-Ito); local to this figure.
\definecolor{oiBlue}{HTML}{0072B2}\definecolor{oiBlueT}{HTML}{DCEBF7}
\definecolor{oiOrange}{HTML}{E69F00}\definecolor{oiOrangeT}{HTML}{FCEFD2}
\definecolor{oiVerm}{HTML}{D55E00}\definecolor{oiVermT}{HTML}{FBE3D6}
\definecolor{oiGray}{HTML}{6B6B6B}\definecolor{oiGrayT}{HTML}{F0F0F0}
\definecolor{oiInk}{HTML}{333333}
% Rounded chip around an instruction name (#1 = fill color, #2 = text).
\def\kchip#1#2{\tikz[baseline=(kc.base)]\node[fill=#1,rounded corners=1.5pt,
  inner xsep=1.4pt,inner ysep=0.9pt,outer sep=0pt] (kc) {#2};}
\begin{tikzpicture}[
  x=1cm,y=1cm,font=\footnotesize,
  nohyph/.style={execute at begin node={\hyphenpenalty=10000\exhyphenpenalty=10000}},
  softshadow/.style={drop shadow={opacity=0.15,shadow xshift=0.45mm,
    shadow yshift=-0.45mm}},
  % Card: tinted fill, accent bar on the left (drawn by \cardframe).
  box/.style={rounded corners=2pt,align=left,inner sep=3pt,outer sep=0pt,
    text width=45mm+2.4pt,nohyph,execute at begin node={\leftskip=2.4pt},
    softshadow},
  heading/.style={font=\footnotesize\bfseries,inner sep=0pt},
  small/.style={align=left,text width=46.2mm,inner sep=0pt,font=\scriptsize,nohyph},
  bubble/.style={draw=oiInk,line width=0.6pt,rounded corners=4pt,align=left,
    inner sep=2.5pt,fill=oiOrangeT,text width=33mm,font=\scriptsize,nohyph,
    softshadow},
  tag/.style={font=\scriptsize,inner sep=1pt},
  side/.style={font=\scriptsize\itshape,inner sep=2pt},
  axislabel/.style={font=\scriptsize,inner sep=1.5pt},
  boundary/.style={line width=1.1pt,draw=oiVerm},
  flow/.style={-{Latex[length=1.8mm]},semithick,draw=oiInk},
]
  % ---------------- Map geometry ----------------
  \def\xL{5.55}\def\xM{9.645}\def\xR{13.74}
  \def\yT{3.70}\def\yM{1.55}\def\yB{-0.62}

  % Accent bar and outline of a card node (#1 = node, #2 = accent color).
  \def\cardframe#1#2{%
    \begin{scope}
      \clip[rounded corners=2pt] (#1.south west) rectangle (#1.north east);
      \fill[#2] (#1.south west) rectangle ([xshift=2.2pt]#1.north west);
    \end{scope}
    \draw[oiInk,line width=0.6pt,rounded corners=2pt]
      (#1.south west) rectangle (#1.north east);}

  % Tail at the lower-left corner of a speech bubble.
  \def\bubbletail#1{%
    \fill[oiOrangeT] ([xshift=3mm,yshift=0.75pt]#1.south west) -- ++(-1.2mm,-1.6mm)
      -- ++(3.0mm,1.6mm) -- cycle;
    \draw[oiInk,line width=0.6pt,line join=round]
      ([xshift=3mm]#1.south west) -- ++(-1.2mm,-1.6mm) -- ++(3.0mm,1.6mm);}

  % ---------------- Left: statement, evidence, instructions ----------------
  \node[heading,anchor=north west] (hhead) at (0,4.34) {Statement $h$};
  \node[box,fill=oiBlueT,anchor=north west] (stmt) at ([yshift=-1.2mm]hhead.south west)
    {``The agent may raise a debit card's ATM limit only after the user gives
     two of four details: date of birth, email, phone number, and address''};
  \cardframe{stmt}{oiBlue}
  \node[small,anchor=north west,text width=46.2mm+2.6pt,
    execute at begin node={\leftskip=2.6pt}] (evid) at ([yshift=-1.4mm]stmt.south west)
    {Supporting reply: ``\ldots your ATM withdrawal request. However,
     I need to first verify your identity \ldots{} please provide me with two
     of the following \ldots? Date of birth, Email address, Phone number,
     Address''};
  \draw[oiBlue,line width=1.1pt,line cap=round]
    ([xshift=0.55pt,yshift=-0.6pt]evid.north west)
    -- ([xshift=0.55pt,yshift=0.6pt]evid.south west);
  \node[heading,anchor=north west] (khead) at ([yshift=-2.6mm]evid.south west)
    {Instructions paired with $h$};
  \node[box,fill=oiOrangeT,anchor=north west] (instr) at ([yshift=-1.2mm]khead.south west)
    {\kchip{oiOrange!45}{$k_1$}\ \ Omit one required input\\
     \kchip{oiOrange!45}{$k_2$}\ \ Repeat a completed request\\
     \textcolor{black!50}{\kchip{oiGrayT}{$k_3$}\ \ Swap source and destination}};
  \cardframe{instr}{oiOrange}
  \node[small,anchor=north west] (na) at ([yshift=-1.2mm]instr.south west)
    {Pair $(h,k_3)$ is not applicable: a limit increase has no source or
     destination.};

  % ---------------- Right: map of limit requests ----------------
  \node[heading,anchor=north west] at (\xL,4.34)
    {Requests to raise the daily ATM limit of the card ending in 4821};
  \path[fill=white,rounded corners=2pt,softshadow] (\xL,\yB) rectangle (\xR,\yT);
  \begin{scope}
    \clip[rounded corners=2pt] (\xL,\yB) rectangle (\xR,\yT);
    \fill[oiVermT] (\xL,\yM) rectangle (\xM,\yT);
  \end{scope}
  \draw[oiGray,line width=0.6pt,rounded corners=2pt] (\xL,\yB) rectangle (\xR,\yT);
  \draw[boundary] (\xM,\yB) -- (\xM,\yT);
  \draw[boundary,dashed] (\xL,\yM) -- (\xR,\yM);

  % Column and row labels.
  \node[axislabel,anchor=south] at ({(\xL+\xM)/2},\yT) {User gives two details};
  \node[axislabel,anchor=south] at ({(\xM+\xR)/2},\yT) {User gives one detail};
  \node[axislabel,anchor=south,rotate=90,align=center] at (\xL,{(\yM+\yT)/2-0.2})
    {First increase\\in 24 hours};
  \node[axislabel,anchor=south,rotate=90,align=center] at (\xL,{(\yB+\yM)/2})
    {Second increase\\in 24 hours};

  % Boundary names.
  \node[font=\scriptsize,anchor=north,inner sep=2pt] at (\xM,\yB) {Boundary of $h$};
  \node[font=\scriptsize,anchor=north west,inner sep=1.5pt] at (\xM+0.08,\yM-0.05)
    {Boundary of one increase per 24 hours};

  % The statement describes the shaded requests.
  \node[font=\footnotesize,anchor=west] (sideh) at ({\xL+0.30},0 |- stmt.east)
    {Requests described by $h$};
  \draw[flow] (stmt.east) -- (sideh.west);

  % Expected behavior on each side.
  \node[side,anchor=south east] at (\xM,\yM) {Agent may raise the limit};
  \node[side,anchor=south east] at (\xR,\yM) {Agent must not raise the limit yet};
  \node[side,anchor=south east] at (\xM,\yB) {Agent must refuse};
  \node[side,anchor=south east] at (\xR,\yB) {Agent must not raise the limit};

  % Transformed tests for pairs: the request from the user in each test.
  \node[bubble,anchor=north west] (t2) at (\xM+0.45,\yT-0.40)
    {``I'm Liang Jinhai. Please raise the ATM limit of my card ending in 4821
     to \$2,000. My date of birth is April 12, 2002.''};
  \node[bubble,anchor=north west] (t1) at (\xL+0.30,\yM-0.40)
    {``Thanks. Please raise my ATM limit again.''};
  \foreach \t in {t1,t2} {\bubbletail{\t}}
  \node[tag,anchor=south west] at ([xshift=1mm]t2.north west) {Transformed test for $(h,k_1)$};
  \node[tag,anchor=south west] at ([xshift=1mm]t1.north west) {Transformed test for $(h,k_2)$};
\end{tikzpicture}
\caption{Combinations (pairs) of a banking statement $h$ with three test transformation instructions $k_1$ to $k_3$.
The boundary of $h$ (solid line) and the boundary of the condition that allows only one increase in 24 hours (dashed line) divide the requests to raise the daily ATM limit of a debit card; $h$ describes the shaded requests, where the agent may raise the limit.
Speech bubbles show the requests from the user in the transformed tests for $(h,k_1)$ and $(h,k_2)$, that is, the tests that the generator can write when it receives $h$ together with $k_1$ or $k_2$; the generator reports $(h,k_3)$ as not applicable.}
\label{fig:motivation}
\end{figure}
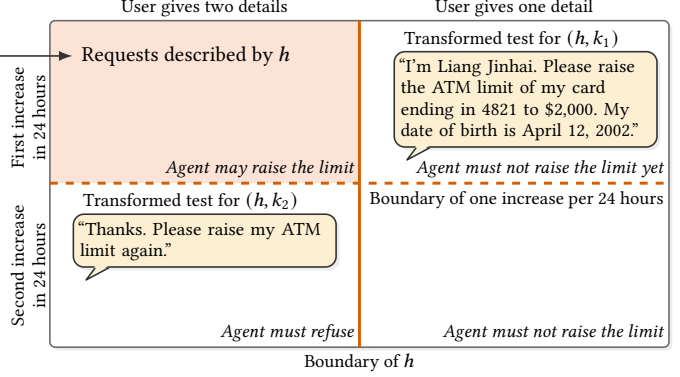

The closest prior work, AgentEval~\cite{agenteval}, explores a black-box agent through conversations in a phase called \emph{discovery} and then generates tests from the resulting records, the \emph{discovery traces}.
For boundary testing, it mines a workflow graph from these traces and prompts an LLM (the \emph{generator}) to write tests that target the graph's nodes and edges by perturbing user messages.
AgentEval's ablation analysis shows that using nodes and edges as explicit targets yields tests that cover more distinct boundaries than prompting an LLM with the traces alone.
This idea of providing explicit targets inspired our work.
However, our analysis of AgentEval's workflow graph and generated tests reveals three problems.
First, each node merges agent replies that reflect the same kind of activity (e.g., processing an exchange), and each edge merges the user messages that move the conversation from one activity to the next.
As a result, a single node or edge may contain several boundaries, so the generator must infer which boundary to test and what behavior to expect.
Because the graph does not specify expected behavior, the generator supplies it.
Many of AgentEval's tests therefore expect behaviors not supported by the agent's \emph{artifacts}: its prompt, tool code, and (if present) the knowledge base it consults.
In our evaluation, 22.0\% to 36.0\% of AgentEval's tests expect behavior that follows only from the domain knowledge of the LLM driving the agent.
In most of these cases, the expected behavior is a general conversational heuristic (e.g., asking for a missing value) that the artifacts do not define.
Second, the size of the graph limits the number of targets.
Third, AgentEval selects all targets before executing any test, so outcomes of earlier tests cannot guide the choice of later targets.

Our method, \tool (Adaptive Test Transformations), keeps the idea of giving the generator explicit targets but replaces AgentEval's nodes and edges with \emph{statements}: natural-language hypotheses that describe a single policy condition and the behavior expected of the agent when that condition holds.
With a statement as its target, the generator is given the expected behavior directly, rather than having to infer it.
An LLM extracts statements from the agent's replies in the discovery traces and stores each statement together with its supporting reply as \emph{evidence}.
Extraction is possible because agents often explicitly state the conditions they apply, for example, when explaining why they refuse a request.

\tool generates \emph{statement-guided tests}, for which the generator receives a statement as its target, in two forms.
For a \emph{plain test}, the generator receives the statement alone and writes a test for the side of the boundary that the statement describes.
For a \emph{transformed test}, the generator also receives a \emph{test transformation instruction}, or simply \emph{instruction}, such as an instruction to change a value in a user request, the wording of the request, or the state of the conversation.
The instruction transforms the test that the generator would write for the statement alone and can thereby move it to the other side of the statement's boundary or to a boundary that the statement does not describe.
Figure~\ref{fig:motivation} shows both cases for a banking-agent statement.

However, exhaustively exploring all possible combinations of statements and instructions (\emph{pairs}) is infeasible.
In one of our experiments, for example, 126 statements and 91 instructions form 11{,}466 pairs, but one run of test generation may try only 400 of them.
Moreover, many pairs are \emph{not applicable}: their instruction changes nothing in a test of their statement (as in the pair $(h,k_3)$ in Figure~\ref{fig:motivation}), and the generator reports them as such.
To choose pairs under these constraints, we introduce \emph{adaptive pair selection}, in which a \emph{selector} chooses first the statement and then the instruction of each pair.
The selector chooses the statement by \emph{novelty order}, which prefers the statements least similar to those that already have a transformed test.
It chooses the instruction for this statement with the bandit algorithm Bayes-UCB~\cite{kaufmann2012bayesucb}, treating the instructions as the arms of a multi-armed bandit problem~\cite{lai1985asymptotically,lattimore2020bandit}.
The selector learns which instructions to choose from two signals that become known only after a pair is tried: whether the generator writes a transformed test (whether the pair is applicable), and the verdict of an LLM \emph{judge} on the executed test (pass, fail, or inconclusive).

To measure \tool's effectiveness, we built a benchmark for boundary tests.
A boundary can come from two sources: it can be defined by the agent's artifacts, or it can be inferred by the LLM that drives the agent from domain knowledge.
For example, an airline booking agent may refuse to book a flight for a past date based on domain knowledge, even when the agent's policy does not include this condition.
We call a boundary \emph{explicit} when it is defined by the artifacts.

In our benchmark, an LLM \emph{auditor} reads the artifacts---which the tester cannot see---and labels each test.
A test is \emph{valid} if the facts it assumes hold in the initial state of the agent's database, its messages lead the conversation to the situation it checks, and every behavior it expects of the agent follows from the artifacts or is necessarily implied by domain knowledge.
A valid test that covers at least one boundary is a \emph{valid boundary test}. It is \emph{explicitly valid} if every expected behavior follows from the artifacts.
The auditor counts the distinct explicit boundaries covered by the valid boundary tests, counting the two sides of each boundary separately.

We evaluate \tool on four domains of $\tau^3$-bench and compare it with AgentEval, which generates tests from the same discovery traces using the same generator model.
Across the four domains, 74.3\% to 88.3\% of \tool's tests are valid boundary tests, and 62.7\% to 83.3\% are explicitly valid.
The transformed tests add 13 to 46 explicit boundaries beyond those covered by the plain tests.
In every domain, \tool produces a higher proportion of explicitly valid tests than AgentEval, and its test suite covers at least as many explicit boundaries and more sides.
As regression tests, \tool's test suites detect all eight seeded faults in the airline domain and four of eight in the retail domain, whereas AgentEval's test suites detect five and two, respectively.

This paper makes three contributions.
\begin{enumerate}
\item \textbf{Statements and test transformation instructions.}
We introduce \emph{statements} to guide test generation.
Each statement is extracted from the discovery traces and gives the generator the behavior expected of the agent under a condition.
We also introduce \emph{test transformation instructions} that transform a statement's test so it can cover the other side of the statement's boundary or another boundary (Sections~\ref{sec:method-statements} and~\ref{sec:method-instructions}).
\item \textbf{Adaptive pair selection.}
We propose a method that adaptively selects statement--instruction pairs.
It selects each pair's statement by novelty and its instruction using the multi-armed bandit algorithm Bayes-UCB. Its feedback consists of two signals that a black-box tester observes: whether the pair is applicable and the judge's verdict on the transformed test (Section~\ref{sec:method-selector}).
\item \textbf{A benchmark for boundary tests.}
We design a benchmark for boundary tests that distinguishes the two sides of each boundary and separates boundaries defined by the agent's artifacts from those inferred by the LLM from domain knowledge (Section~\ref{sec:eval-assessment}).
\end{enumerate}

% ======================================================================
% 2 Preliminaries
% ======================================================================
\section{Preliminaries}
\label{sec:prelim}

\subsection{Discovery, Generation, and Execution}
\label{sec:prelim-framework}

Our testing framework builds on AgentEval~\cite{agenteval} and consists of the three phases shown in Figure~\ref{fig:framework}.
In \emph{discovery}, the tester explores the agent through conversations; \emph{test generation} writes the tests; and \emph{execution} runs and judges them.
Our discovery and execution phases match AgentEval's, and we focus on improving test generation.

% Figure 2 (fig:framework): the testing framework in two tiers.
% Top tier: the tester with its three phases (discovery, test generation,
% execution). Bottom tier: the agent under test as a black box. The dashed
% line between the tiers is the interface; the LLM driver (discovery) and the
% runner (execution) cross it with the operations reset and invoke only.
% Colored after the shared style guide (Okabe-Ito): discovery, agent and
% traces gray; test generation purple; execution, judge and verdicts green;
% the verdict feedback of AdaT^2 dashed blue. Colors are defined inside the
% figure and the tikz library shadows is loaded here, so the preamble needs no
% change.
\usetikzlibrary{shadows}%
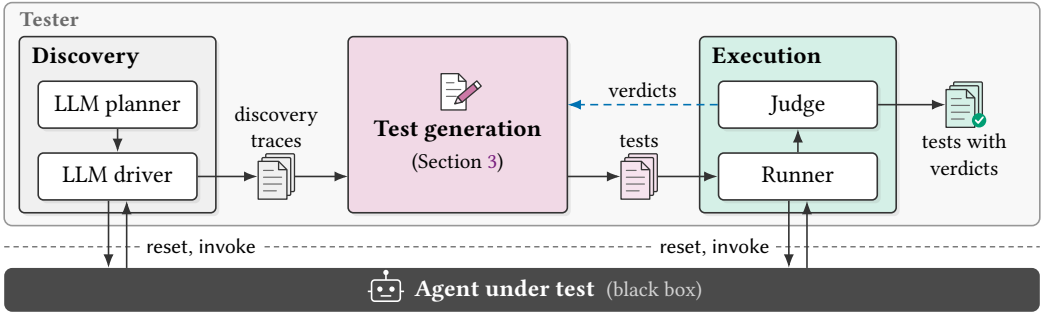
\begin{figure}[t]
\centering
\definecolor{fwInk}{HTML}{333333}%
\definecolor{fwGray}{HTML}{6B6B6B}%
\definecolor{fwGrayT}{HTML}{F0F0F0}%
\definecolor{fwPurple}{HTML}{CC79A7}%
\definecolor{fwPurpleT}{HTML}{F6E3EE}%
\definecolor{fwGreen}{HTML}{009E73}%
\definecolor{fwGreenT}{HTML}{D5F0E7}%
\definecolor{fwBlue}{HTML}{0072B2}%
\definecolor{fwDark}{HTML}{4A4A4A}%
\begin{tikzpicture}[
  x=1cm,y=1cm,font=\small,
  softshadow/.style={drop shadow={opacity=0.15,shadow xshift=0.45mm,
    shadow yshift=-0.45mm}},
  card/.style={draw=fwInk,line width=0.6pt,rounded corners=2pt},
  comp/.style={draw=fwInk,line width=0.6pt,rounded corners=2pt,fill=white,
    align=center,minimum width=21mm,minimum height=6.2mm,inner sep=2pt,
    softshadow,execute at begin node=\strut},  % same text height in every box
  phasename/.style={font=\small\bfseries,anchor=north west,inner sep=0pt},
  note/.style={font=\footnotesize,inner sep=1pt,align=center},
  flow/.style={-{Latex[length=1.8mm]},semithick,draw=fwInk},
  feedback/.style={-{Latex[length=1.8mm]},semithick,draw=fwBlue,
    dash pattern=on 3pt off 2pt},
]
  % ---------------- Geometry ----------------
  \def\W{13.7}                  % width of both tiers
  \def\yU{-1.35}\def\yL{-2.30}  % upper and lower component rows
  \def\cT{-0.45}\def\cB{-2.78}  % top and bottom of the phase cards
  \def\bB{-2.95}                % bottom of the tester tier
  \def\yI{-3.23}                % interface line
  \def\aT{-3.51}\def\aB{-4.09}  % agent tier

  % Stack of three pages centered at #2, front page filled with #3 (#1 = name
  % of an invisible node covering the stack, for attaching arrows).
  \def\docstack#1#2#3{%
    \begin{scope}[shift={#2},line width=0.5pt,draw=fwInk]
      \foreach \o in {0.10,0.05}
        \draw[fill=#3,rounded corners=0.5pt] (-0.24+\o,-0.30+\o) rectangle (0.14+\o,0.20+\o);
      \draw[fill=#3,line join=round] (-0.24,-0.30) -- (0.14,-0.30) -- (0.14,0.09)
        -- (0.03,0.20) -- (-0.24,0.20) -- cycle;
      \draw[line join=round] (0.03,0.20) -- (0.03,0.09) -- (0.14,0.09);
      \draw[fwGray,line width=0.45pt,line cap=round]
        (-0.17,0.08) -- (-0.04,0.08) (-0.17,-0.01) -- (0.07,-0.01)
        (-0.17,-0.10) -- (0.07,-0.10) (-0.17,-0.19) -- (0.00,-0.19);
    \end{scope}
    \node[inner sep=0pt,minimum width=4.8mm,minimum height=6mm] (#1) at #2 {};}

  % ---------------- Tester tier ----------------
  \draw[fill=fwGrayT!45,draw=fwGray!70,line width=0.6pt,rounded corners=3pt]
    (0,0) rectangle (\W,\bB);
  \node[font=\footnotesize\bfseries,text=fwGray,anchor=north west,inner sep=0pt]
    at (0.20,-0.12) {Tester};

  % Phase cards.
  \draw[card,fill=fwGrayT,softshadow] (0.20,\cT) rectangle (2.80,\cB);
  % Same 0.6pt ink outline as the other cards; size and fill mark it.
  \draw[card,fill=fwPurple!28!white,softshadow]
    (4.55,\cT) rectangle (7.45,\cB);
  \draw[card,fill=fwGreenT,softshadow] (9.20,\cT) rectangle (11.80,\cB);
  \node[phasename] at (0.36,\cT-0.13) {Discovery};
  \node[phasename] at (9.36,\cT-0.13) {Execution};

  % Discovery: an LLM planner and an LLM driver.
  \node[comp] (planner) at (1.50,\yU) {LLM planner};
  \node[comp] (driver) at (1.50,\yL) {LLM driver};
  \draw[flow] (planner) -- (driver);

  % Test generation (Section 3), the contribution.
  \begin{scope}[shift={(5.98,-1.17)},line width=0.5pt,draw=fwInk]
    % Glyph: a page and a pencil.
    \draw[fill=white,line join=round] (-0.20,-0.25) -- (0.13,-0.25) -- (0.13,0.13)
      -- (0.03,0.23) -- (-0.20,0.23) -- cycle;
    \draw[line join=round] (0.03,0.23) -- (0.03,0.13) -- (0.13,0.13);
    \draw[fwGray,line width=0.45pt,line cap=round]
      (-0.13,0.11) -- (-0.03,0.11) (-0.13,0.02) -- (0.06,0.02) (-0.13,-0.07) -- (0.00,-0.07);
    \begin{scope}[shift={(-0.02,-0.17)},rotate=40,line join=round]
      \draw[fill=fwPurpleT] (0,0) -- (0.09,0.045) -- (0.09,-0.045) -- cycle;
      \fill[fwInk] (0,0) -- (0.03,0.015) -- (0.03,-0.015) -- cycle;
      \draw[fill=fwPurple] (0.09,-0.045) rectangle (0.36,0.045);
      \draw[fill=white] (0.36,-0.045) rectangle (0.42,0.045);
    \end{scope}
  \end{scope}
  \node[align=center,inner sep=0pt,anchor=north] at (6.00,-1.55)
    {\textbf{Test generation}\\[1pt]{\footnotesize (Section~\ref{sec:method})}};

  % Execution: a runner and a judge.
  \node[comp] (runner) at (10.50,\yL) {Runner};
  \node[comp] (judge) at (10.50,\yU) {Judge};
  \draw[flow] (runner) -- (judge);

  % Discovery traces into test generation, tests into execution.
  \docstack{traces}{(3.60,\yL)}{fwGrayT}
  \node[note,anchor=south] at ([yshift=1pt]traces.north) {discovery\\traces};
  \draw[flow] (driver.east) -- (traces.west);
  \draw[flow] (traces.east) -- (4.55,\yL);
  \docstack{tests}{(8.40,\yL)}{fwPurpleT}
  \node[note,anchor=south] at ([yshift=1pt]tests.north) {tests};
  \draw[flow] (7.45,\yL) -- (tests.west);
  \draw[flow] (tests.east) -- (runner.west);

  % Verdicts: fed back to test generation (AdaT^2) and the output.
  \draw[feedback] (judge.west) -- node[note,above=1.5pt] {verdicts} (7.45,\yU);
  \docstack{out}{(12.70,\yU)}{fwGreenT}
  \node[note,anchor=north] at ([yshift=-1pt]out.south) {tests with\\verdicts};
  \draw[flow] (judge.east) -- (out.west);
  \begin{scope}[shift={([xshift=-0.5mm,yshift=1mm]out.south east)}]
    % Verdict badge.
    \fill[fwGreen] (0,0) circle (0.095);
    \draw[white,line width=0.7pt,line cap=round,line join=round]
      (-0.045,0.0) -- (-0.012,-0.035) -- (0.05,0.04);
  \end{scope}

  % ---------------- Interface ----------------
  \draw[fwGray,line width=0.5pt,dash pattern=on 2pt off 1.5pt] (0,\yI) -- (\W,\yI);
  \foreach \c/\s in {driver/1,runner/-1} {
    \draw[flow] ([xshift=-1.2mm]\c.south) -- ([xshift=-1.2mm]\c.south |- 0,\aT);
    \draw[flow] ([xshift=1.2mm]\c.south |- 0,\aT) -- ([xshift=1.2mm]\c.south);}
  \node[note,fill=white,anchor=west,inner xsep=2pt] at (1.80,\yI)
    {$\mathsf{reset}$, $\mathsf{invoke}$};
  \node[note,fill=white,anchor=east,inner xsep=2pt] at (10.20,\yI)
    {$\mathsf{reset}$, $\mathsf{invoke}$};

  % ---------------- Agent tier (black box) ----------------
  \fill[fwDark,rounded corners=3pt] (0,\aT) rectangle (\W,\aB);
  \begin{scope}[shift={(5.05,{(\aT+\aB)/2})},white,line width=0.6pt]
    % Robot glyph.
    \draw[rounded corners=1.2pt] (-0.19,-0.15) rectangle (0.19,0.13);
    \draw (0,0.13) -- (0,0.21);
    \fill (0,0.23) circle (0.035);
    \fill (-0.08,0.01) circle (0.035) (0.08,0.01) circle (0.035);
    \draw (-0.07,-0.08) -- (0.07,-0.08);
    \draw (-0.19,-0.02) -- (-0.24,-0.02) (0.19,-0.02) -- (0.24,-0.02);
  \end{scope}
  \node[anchor=west,text=white,inner sep=0pt] at (5.42,{(\aT+\aB)/2})
    {\textbf{Agent under test}\ \ \textcolor{white!75!fwDark}{\footnotesize(black box)}};
\end{tikzpicture}
\caption{The testing framework.
Discovery explores the agent in conversation, test generation writes tests from the discovery traces (the recorded conversations of discovery), and execution runs each test on the agent and judges it.
In \tool, the selector of test generation uses the judge's verdicts on the executed tests as a signal for choosing pairs (dashed arrow, Section~\ref{sec:method-selector}).}
\label{fig:framework}
\end{figure}

\begin{figure}[t]
\centering
\begin{tikzpicture}[
  x=1cm,y=1cm,font=\small,
  operation/.style={draw,rounded corners=1.5pt,align=center,
    inner sep=2.5pt,minimum height=6mm,text width=15mm,fill=white,
    execute at begin node=\strut},
  artifact/.style={draw,align=center,inner sep=2.5pt,minimum height=6mm,fill=white},
  flow/.style={-{Latex[length=1.8mm]},semithick},
  message/.style={flow},
  reply/.style={flow,dashed},
  lifeline/.style={thin,densely dotted},
  activation/.style={draw,thin,fill=white,minimum width=1.6mm,inner sep=0pt},
  turnbrace/.style={decorate,decoration={brace,amplitude=3.5pt,mirror},thin},
  note/.style={font=\footnotesize,inner sep=1pt,align=center},
  label text/.style={inner sep=2pt}]
  % Lifeline x positions, right edge; event y positions.
  \def\xT{3.85}\def\xA{9.75}\def\xR{10.75}
  \def\xF{1.75}\def\yF{-0.55}\def\yB{-3.85}
  \def\yReset{-1.05}
  \def\yIa{-1.65}\def\yRa{-2.3}
  \def\yIb{-2.95}\def\yRb{-3.6}

  % Session frame with its tab.
  \draw (\xF,\yF) rectangle (\xR,\yB);
  \node[note,anchor=north west,align=left,inner sep=2.5pt] (tab) at (\xF,\yF)
    {Session\\(turn limit 2)};
  \draw (tab.south west) -- ([xshift=-1.5mm]tab.south east)
    -- ([yshift=1.5mm]tab.south east) -- (tab.north east);

  % Participants and lifelines.
  \node[operation] (tester) at (\xT,0) {Tester};
  \node[operation] (agent) at (\xA,0) {Agent};
  \draw[lifeline] (tester.south) -- (\xT,\yB+0.1);
  \draw[lifeline] (agent.south) -- (\xA,\yB+0.1);

  % reset starts a fresh session.
  \draw[message] (\xT,\yReset) -- node[above,label text] {$\mathsf{reset}$: start a fresh session} (\xA,\yReset);

  % Turn 1.
  \node[activation,anchor=north,minimum height={(\yIa-\yRa)*1cm}] (act1) at (\xA,\yIa) {};
  \draw[message] (\xT,\yIa) -- node[above,label text] {$\mathsf{invoke}$: ``Look up reservation NO6JO3''} (act1.north west);
  \draw[reply] (act1.south west) -- node[above,label text] {``Two unflown business-class flights''} (\xT,\yRa);

  % Turn 2.
  \node[activation,anchor=north,minimum height={(\yIb-\yRb)*1cm}] (act2) at (\xA,\yIb) {};
  \draw[message] (\xT,\yIb) -- node[above,label text] {$\mathsf{invoke}$: ``Economy for flight 1 only?''} (act2.north west);
  \draw[reply] (act2.south west) -- node[above,label text] {``Not allowed: cabins must match''} (\xT,\yRb);

  % Turns and the trace that records them in order.
  \draw[turnbrace] (\xT-0.12,\yIa+0.06) -- coordinate[midway] (b1) (\xT-0.12,\yRa-0.06);
  \draw[turnbrace] (\xT-0.12,\yIb+0.06) -- coordinate[midway] (b2) (\xT-0.12,\yRb-0.06);
  \def\yTop{-1.325}\def\yMid{-2.625}
  \draw[fill=white] (0.05,\yTop) rectangle (1.45,\yB);
  \draw (0.05,\yMid) -- (1.45,\yMid);
  \node[note,anchor=south,inner sep=2pt] at (0.75,\yTop) {Trace};
  \node at (0.75,{(\yTop+\yMid)/2}) {turn 1};
  \node at (0.75,{(\yMid+\yB)/2}) {turn 2};
  \draw[flow] ([xshift=-4pt]b1) -- (1.45,0 |- b1);
  \draw[flow] ([xshift=-4pt]b2) -- (1.45,0 |- b2);
\end{tikzpicture}
\caption{Interaction between the tester and the airline agent of $\tau^3$-bench in a session with a turn limit of two, in which the tester sends the step hints of the test in Figure~\ref{fig:test}, shortened.
The trace records each user message and the agent's reply to it as one turn, in order.}
\label{fig:interaction}
\end{figure}
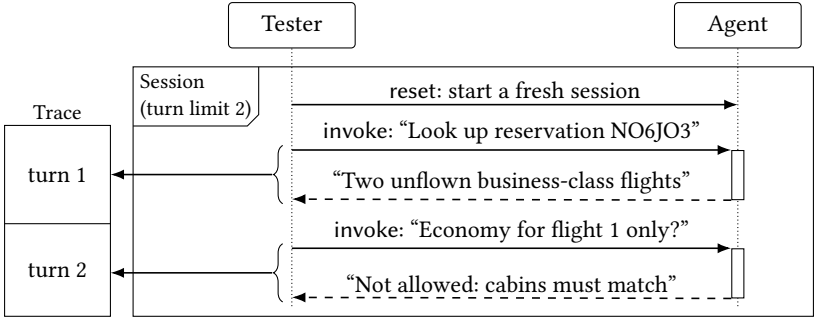

\paragraph{Interaction.}
The tester interacts with the agent through two operations (Figure~\ref{fig:interaction}).
The operation $\mathsf{reset}$ starts a fresh \emph{session}, a conversation that begins in the initial \emph{agent state}, such as the database that the agent's tools read and change.
The operation $\mathsf{invoke}$ sends the agent one \emph{user message}, which the tester writes as the user, and returns the agent's reply.
A user message and the agent's reply form a \emph{turn}; a \emph{trace} records the turns of a session, and a \emph{turn limit} bounds the number of turns.
The tester observes only the user messages and the agent's replies.

\paragraph{Discovery.}
In discovery, the tester runs a fixed number of sessions to learn which tasks the agent supports, what inputs they require, and how the agent responds to user requests.
Each session has an \emph{objective} (what the tester wants to learn in the session) and a \emph{strategy} (how the tester pursues the objective through conversation).
Before each session, an LLM planner chooses the objective and the strategy based on the traces of the earlier sessions.
During the session, an LLM driver writes each user message, and the session ends when the driver decides that the objective is met or cannot be met, or when the session reaches the turn limit.
The traces of these sessions, the \emph{discovery traces}, are the input for test generation.

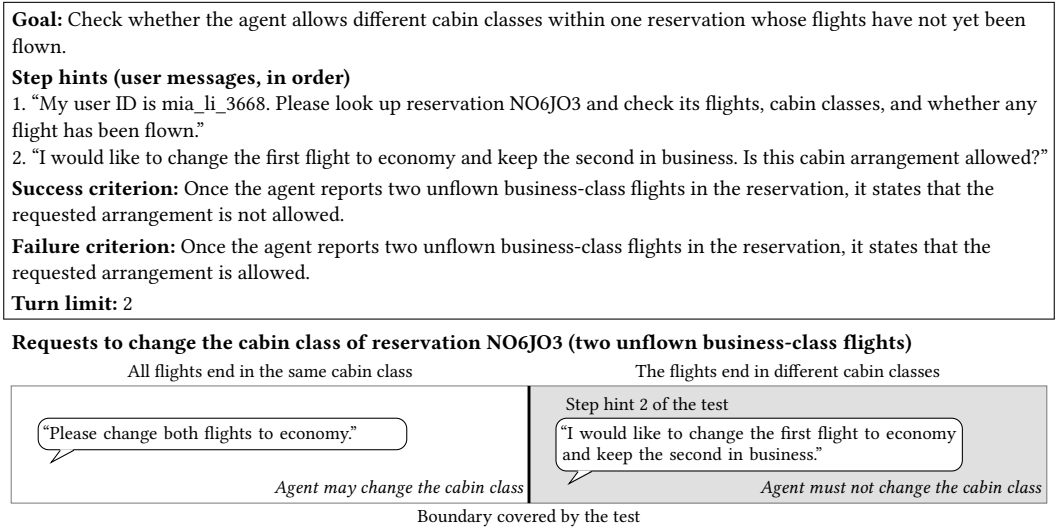
\begin{figure}[t]
\centering
\fbox{%
\begin{minipage}{\dimexpr\columnwidth-2\fboxsep-2\fboxrule\relax}
\footnotesize\raggedright
\textbf{Goal:} Check whether the agent allows different cabin classes within one reservation whose flights have not yet been flown.\par
\smallskip
\textbf{Step hints (user messages, in order)}\par
1. ``My user ID is mia\_li\_3668.
Please look up reservation NO6JO3 and check its flights, cabin classes, and whether any flight has been flown.''\par
2. ``I would like to change the first flight to economy and keep the second in business.
Is this cabin arrangement allowed?''\par
\smallskip
\textbf{Success criterion:} Once the agent reports two unflown business-class flights in the reservation, it states that the requested arrangement is not allowed.\par
\smallskip
\textbf{Failure criterion:} Once the agent reports two unflown business-class flights in the reservation, it states that the requested arrangement is allowed.\par
\smallskip
\textbf{Turn limit:} 2
\end{minipage}%
}\par
\medskip
\begin{tikzpicture}[
  x=1cm,y=1cm,font=\footnotesize,
  nohyph/.style={execute at begin node={\hyphenpenalty=10000\exhyphenpenalty=10000}},
  heading/.style={font=\footnotesize\bfseries,inner sep=0pt},
  bubble/.style={draw,rounded corners=4pt,align=left,inner sep=2.5pt,fill=white,
    text width=52mm,font=\scriptsize,nohyph},
  tag/.style={font=\scriptsize,inner sep=1pt},
  side/.style={font=\scriptsize\itshape,inner sep=2pt},
  axislabel/.style={font=\scriptsize,inner sep=1.5pt},
  boundary/.style={line width=1.1pt},
]
  \def\xL{0}\def\xM{6.85}\def\xR{13.7}
  \def\yT{1.55}\def\yB{0}
  \def\bubbletail#1{%
    \fill[white] ([xshift=3mm,yshift=0.4pt]#1.south west) -- ++(-1.2mm,-1.6mm)
      -- ++(3.0mm,1.6mm) -- cycle;
    \draw ([xshift=3mm]#1.south west) -- ++(-1.2mm,-1.6mm) -- ++(3.0mm,1.6mm);}

  \node[heading,anchor=south west] at (\xL,\yT+0.42)
    {Requests to change the cabin class of reservation NO6JO3 (two unflown business-class flights)};
  \fill[black!12] (\xM,\yB) rectangle (\xR,\yT);
  \draw[black!50] (\xL,\yB) rectangle (\xR,\yT);
  \draw[boundary] (\xM,\yB) -- (\xM,\yT);

  \node[axislabel,anchor=south] at ({(\xL+\xM)/2},\yT) {All flights end in the same cabin class};
  \node[axislabel,anchor=south] at ({(\xM+\xR)/2},\yT) {The flights end in different cabin classes};
  \node[font=\scriptsize,anchor=north,inner sep=2pt] at (\xM,\yB) {Boundary covered by the test};

  \node[side,anchor=south east] at (\xM,\yB) {Agent may change the cabin class};
  \node[side,anchor=south east] at (\xR,\yB) {Agent must not change the cabin class};

  \node[bubble,anchor=north west,text width=47mm] (b1) at (\xL+0.35,\yT-0.42)
    {``Please change both flights to economy.''};
  \node[bubble,anchor=north west] (b2) at (\xM+0.35,\yT-0.42)
    {``I would like to change the first flight to economy and keep the second in business.''};
  \foreach \t in {b1,b2} {\bubbletail{\t}}
  \node[tag,anchor=south west] at ([xshift=1mm]b2.north west) {Step hint 2 of the test};
\end{tikzpicture}
\caption{An illustrative test of the airline agent (top) and the boundary that it covers (bottom).
The user asks to change the cabin class of only one of two unflown business-class flights.
The boundary whose condition is that all flights of a reservation have the same cabin class (solid line) divides the requests to change the cabin class of reservation NO6JO3, and the test covers the shaded side, where the agent must not change the cabin class.}
\label{fig:test}
\end{figure}

\paragraph{Tests.}
Following AgentEval, a \emph{test} contains a \emph{goal}, \emph{step hints}, \emph{criteria}, and a turn limit (Figure~\ref{fig:test}, explained in Section~\ref{sec:prelim-example}).
The goal names the behavior of the agent that the test exercises, and the step hints are the user messages that the tester sends in order.
\emph{Success criteria} describe the agent's expected behavior, and \emph{failure criteria} describe incorrect behavior the tester can observe.

\paragraph{Execution and judgment.}
A \emph{runner} executes a test in a fresh session: it sends the step hints in order until it has sent all of them or the conversation reaches the turn limit, and it records the conversation as the \emph{execution trace}.
An LLM \emph{judge} compares the execution trace with the goal and the criteria of the test and returns a \emph{verdict}.
The verdict is pass when the agent achieves the test goal, and the failure criteria do not hold.
It fails when a failure criterion holds or when the agent clearly fails the goal and shows incorrect behavior that the tester can observe.
Otherwise, the verdict is inconclusive.

\subsection{Boundary Tests}
\label{sec:prelim-boundary}

In program testing, a boundary separates neighboring groups of inputs that a program handles or is specified to handle differently~\cite{clarke1982domain,hierons2006bva}.
The input of a conversational agent consists of the requests from the user.
A \emph{boundary} of an agent is a division of these requests into two groups, its \emph{sides}, that the agent must handle differently.
A \emph{condition} of the agent's policy defines a boundary: the requests that satisfy the condition lie on one side, and the requests that violate it lie on the other.
Whether a request satisfies the condition can depend on the values in the request and on the state in which the user makes it, that is, on what the conversation has established and what the agent state contains.
For example, the same request to raise the daily ATM limit of a debit card lies on different sides of the dashed boundary in Figure~\ref{fig:motivation} before and after the agent has raised the limit within the last 24 hours.
The policy specifies the behavior expected of the agent on one side of a boundary or on both sides.
A boundary is \emph{explicit} when the agent's artifacts, that is, its prompt, the code of its tools, and the knowledge base that it consults, define its condition and the behavior expected on at least one of its sides.

A \emph{boundary test} sends the agent a request from one side of a boundary and checks whether the agent then shows the behavior expected on that side, and we say that the test \emph{covers} this side of the boundary.
A test with several criteria can cover several sides or boundaries.

\subsection{Assumptions}
\label{sec:prelim-assumption}

We make three assumptions.
First, the tester can reset the agent to a fresh session with the same initial agent state, a standard requirement of automated testing.
Second, the testing budget is limited: every attempt to generate a test costs an LLM call, and a written test costs additional LLM calls to execute and judge, so the tester must cover many boundaries in a few attempts.
Third, the agent complies with its policy in most conversations, so its faults are rare and its replies during discovery mostly show the expected behavior.

\subsection{Running Example}
\label{sec:prelim-example}

The running example of this paper is the airline agent of $\tau^3$-bench~\cite{tau3bench}, which helps users book, change, and cancel flight reservations.
The policy in the agent's prompt requires that all flights of a reservation have the same cabin class.
Figure~\ref{fig:test} shows the boundary that this condition defines among the requests to change the cabin class of reservation NO6JO3, which has two unflown business-class flights.
A request to change both flights to economy lies on the side where the agent can change the cabin class.
A request to change only the first flight to economy lies on the other side, where the agent must not change the cabin class, and the test in Figure~\ref{fig:test} covers this side.

% ======================================================================
% 3 Method
% ======================================================================
\section{Method}
\label{sec:method}

\subsection{Overview}
\label{sec:method-overview}

% Figure 5 (fig:architecture): test generation in AdaT^2, "adaptive loop"
% design. Inputs enter from the left: discovery traces -> statement
% extraction -> statements H, and the instruction catalog K. Upper lane:
% plain tests, one straight pass (statement h -> generator -> execution and
% judgment). Lower lane: transformed tests, drawn as a loop of attempts on a
% soft track (adaptive pair selection -> pair (h,k) -> generator -> execution
% and judgment; dashed blue feedback "applicable?" from the generator and
% "verdict" from execution and judgment back to selection). Both lanes join
% into the tests with verdicts. Data (traces, statements, instructions, tests
% with verdicts) are drawn as the stack glyphs of Figure 2. Style of Figures 1
% and 2 (Okabe-Ito tints, 0.6pt #333333 rounded outlines, soft shadows on
% rounded boxes only): statements blue, instructions orange, generator purple,
% execution/judgment/verdicts green, selection sky, feedback dashed blue,
% traces gray. Colors are defined inside the figure and the tikz library
% shadows is loaded here, so the preamble needs no change.
\usetikzlibrary{shadows,bending}%
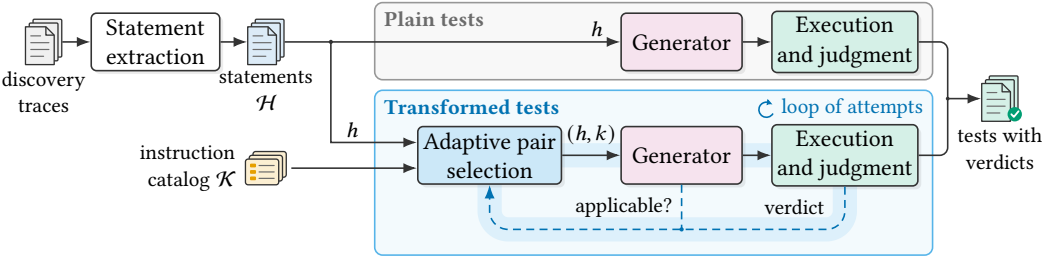
\begin{figure}[t]
\centering
\definecolor{arInk}{HTML}{333333}%
\definecolor{arGray}{HTML}{6B6B6B}%
\definecolor{arGrayT}{HTML}{F0F0F0}%
\definecolor{arBlue}{HTML}{0072B2}%
\definecolor{arBlueT}{HTML}{DCEBF7}%
\definecolor{arOrange}{HTML}{E69F00}%
\definecolor{arOrangeT}{HTML}{FCEFD2}%
\definecolor{arSky}{HTML}{56B4E9}%
\definecolor{arPurple}{HTML}{CC79A7}%
\definecolor{arPurpleT}{HTML}{F6E3EE}%
\definecolor{arGreen}{HTML}{009E73}%
\definecolor{arGreenT}{HTML}{D5F0E7}%
\begin{tikzpicture}[
  x=1cm,y=1cm,font=\small,
  softshadow/.style={drop shadow={opacity=0.15,shadow xshift=0.45mm,
    shadow yshift=-0.45mm}},
  comp/.style={draw=arInk,line width=0.6pt,rounded corners=2pt,fill=white,
    align=center,minimum height=7.6mm,inner sep=2pt,softshadow},
  lane/.style={line width=0.6pt,rounded corners=3pt},
  lanename/.style={font=\footnotesize\bfseries,anchor=north west,inner sep=0pt},
  note/.style={font=\footnotesize,inner sep=1pt,align=center},
  flow/.style={-{Latex[length=1.8mm]},semithick,draw=arInk},
  feedback/.style={-{Latex[length=1.8mm]},semithick,draw=arBlue,
    dash pattern=on 3pt off 2pt},
]
  % ---------------- Geometry ----------------
  \def\yA{0}       % upper lane (plain tests), also the row of the statements
  \def\yB{-1.50}   % lower lane (transformed tests)
  \def\yR{-2.48}   % return path of the loop
  \def\xJ{12.44}   % join of the tests with verdicts
  \def\xS{6.38}\def\xG{8.92}\def\xE{11.08}  % selection, generator, execution
  \def\lL{4.86}\def\lR{12.24}               % left and right ends of the lanes
  \def\xF{4.28}    % fork of the statements

  % Stack of three pages centered at #2, front page filled with #3 (#1 = name
  % of an invisible node covering the stack, for attaching arrows).
  \def\docstack#1#2#3{%
    \begin{scope}[shift={#2},line width=0.5pt,draw=arInk]
      \foreach \o in {0.10,0.05}
        \draw[fill=#3,rounded corners=0.5pt] (-0.24+\o,-0.30+\o) rectangle (0.14+\o,0.20+\o);
      \draw[fill=#3,line join=round] (-0.24,-0.30) -- (0.14,-0.30) -- (0.14,0.09)
        -- (0.03,0.20) -- (-0.24,0.20) -- cycle;
      \draw[line join=round] (0.03,0.20) -- (0.03,0.09) -- (0.14,0.09);
      \draw[arGray,line width=0.45pt,line cap=round]
        (-0.17,0.08) -- (-0.04,0.08) (-0.17,-0.01) -- (0.07,-0.01)
        (-0.17,-0.10) -- (0.07,-0.10) (-0.17,-0.19) -- (0.00,-0.19);
    \end{scope}
    \node[inner sep=0pt,minimum width=4.8mm,minimum height=6mm] (#1) at #2 {};}

  % Catalog of instruction cards centered at #2 (#1 = covering node).
  \def\catalog#1#2{%
    \begin{scope}[shift={#2},line width=0.5pt,draw=arInk]
      \foreach \o in {0.10,0.05}
        \draw[fill=arOrangeT,rounded corners=1pt] (-0.27+\o,-0.22+\o) rectangle (0.17+\o,0.12+\o);
      \draw[fill=arOrangeT,rounded corners=1pt] (-0.27,-0.22) rectangle (0.17,0.12);
      \foreach \y in {0.035,-0.05,-0.135} {
        \fill[arOrange,rounded corners=0.4pt] (-0.21,\y-0.025) rectangle (-0.13,\y+0.025);
        \draw[arGray,line width=0.45pt,line cap=round] (-0.09,\y) -- (0.11,\y);}
    \end{scope}
    \node[inner sep=0pt,minimum width=5.4mm,minimum height=4.4mm] (#1) at #2 {};}

  % ---------------- Lanes ----------------
  \draw[lane,fill=arGrayT!45,draw=arGray!70] (\lL,0.52) rectangle (\lR,-0.50);
  \node[lanename,text=arGray] at (\lL+0.12,0.42) {Plain tests};
  \draw[lane,fill=arSky!7,draw=arSky] (\lL,-0.64) rectangle (\lR,-2.82);
  \node[lanename,text=arBlue] at (\lL+0.12,-0.74) {Transformed tests};
  % The loop of attempts: a soft track under selection, generator,
  % execution and judgment, and the feedback.
  \draw[arSky!20,line width=3.2mm,rounded corners=7pt,line join=round]
    (\xS,\yB) rectangle (\xE,\yR);

  % ---------------- Inputs: statements and instructions ----------------
  \docstack{traces}{(0.48,\yA)}{arGrayT}
  \node[note,anchor=north] at ([yshift=-1pt]traces.south) {discovery\\traces};
  \node[comp,minimum width=17mm] (extract) at (1.96,\yA) {Statement\\extraction};
  \docstack{H}{(3.42,\yA)}{arBlueT}
  \node[note,anchor=north] at ([yshift=-1pt]H.south) {statements\\$\mathcal H$};
  \catalog{K}{(3.42,{\yB-0.17})}
  \node[note,anchor=east,align=right] at ([xshift=-2.5pt]K.west)
    {instruction\\catalog $\mathcal K$};
  \draw[flow] (traces.east) -- (extract.west);
  \draw[flow] (extract.east) -- (H.west);
  % Fork: each statement h goes to both lanes.
  \draw[semithick,draw=arInk] (H.east) -- (\xF,\yA);
  \fill[arInk] (\xF,\yA) circle (0.8pt);

  % ---------------- Upper lane: plain tests, one pass ----------------
  \node[comp,fill=arPurpleT,minimum width=16mm] (genP) at (\xG,\yA) {Generator};
  \node[comp,fill=arGreenT,minimum width=19mm] (execP) at (\xE,\yA) {Execution\\and judgment};
  \draw[flow] (\xF,\yA) -- (genP.west);
  \node[note,anchor=south] at (7.80,\yA+0.03) {$h$};
  \draw[flow] (genP) -- (execP);

  % ---------------- Lower lane: transformed tests, a loop of attempts ----------------
  \node[comp,fill=arSky!30,minimum width=19mm] (select) at (\xS,\yB) {Adaptive pair\\selection};
  \node[comp,fill=arPurpleT,minimum width=16mm] (genT) at (\xG,\yB) {Generator};
  \node[comp,fill=arGreenT,minimum width=19mm] (execT) at (\xE,\yB) {Execution\\and judgment};
  \draw[flow] (\xF,\yA) |- ([yshift=1.7mm]select.west);
  \node[note,anchor=south] at ({(\xF+\lL)/2},{\yB+0.17+0.03}) {$h$};
  \draw[flow] ([xshift=0.6mm]K.east |- 0,{\yB-0.17}) -- ([yshift=-1.7mm]select.west);
  \draw[flow] (select) -- node[note,above=3pt] {$(h,k)$} (genT);
  \draw[flow] (genT) -- (execT);

  % Feedback: applicability from the generator, verdict from the judge.
  \draw[feedback,rounded corners=7pt] (execT.south) -- (\xE,\yR) -- (\xS,\yR)
    -- (select.south);
  \draw[feedback,-] (genT.south) -- (\xG,\yR);
  \fill[arBlue] (\xG,\yR) circle (0.8pt);
  \node[note,anchor=east] at (\xG-0.08,{(\yR+\yB-0.38)/2}) {applicable?};
  \node[note,anchor=east] at (\xE-0.23,{(\yR+\yB-0.38)/2}) {verdict};
  \node[note,anchor=north east,text=arBlue,inner sep=0pt] (loopname) at (\lR-0.12,-0.74)
    {loop of attempts};
  \begin{scope}[shift={([xshift=-1.9mm,yshift=-0.2mm]loopname.west)}]
    \draw[arBlue,line width=0.6pt,-{Latex[length=0.8mm,width=0.9mm,bend]}]
      (-30:0.10) arc (-30:-320:0.10);
  \end{scope}

  % ---------------- Output: tests with verdicts ----------------
  \def\yO{-0.75}
  \docstack{out}{(13.13,\yO)}{arGreenT}
  \draw[semithick,draw=arInk,rounded corners=3pt] (execP.east) -| (\xJ,\yO);
  \draw[semithick,draw=arInk,rounded corners=3pt] (execT.east) -| (\xJ,\yO);
  \draw[flow] (\xJ,\yO) -- (out.west);
  \fill[arInk] (\xJ,\yO) circle (0.7pt);
  \node[note,anchor=north] at ([yshift=-1pt]out.south) {tests with\\verdicts};
  \begin{scope}[shift={([xshift=-0.5mm,yshift=1mm]out.south east)}]
    % Verdict badge.
    \fill[arGreen] (0,0) circle (0.095);
    \draw[white,line width=0.7pt,line cap=round,line join=round]
      (-0.045,0.0) -- (-0.012,-0.035) -- (0.05,0.04);
  \end{scope}
\end{tikzpicture}
\caption{Test generation in \tool.
Statements extracted from the discovery traces guide two ways of generating tests.
For a plain test, the generator receives a statement alone.
For a transformed test, adaptive pair selection combines a statement with a test transformation instruction, and the generator receives this pair.
Whether the generator finds the pair applicable and the judge's verdict on the transformed test (dashed arrows) guide later choices of adaptive pair selection.}
\label{fig:architecture}
\end{figure}

Figure~\ref{fig:architecture} shows how \tool turns discovery traces into tests.
It extracts statements from the traces, writes plain tests for them, and writes transformed tests in a loop in which the outcome of each attempt guides the choice of the next pair.

\begin{figure}[t]
\centering
\begin{tikzpicture}[
  x=1cm,y=1cm,font=\small,
  operation/.style={draw,rounded corners=1.5pt,align=center,
    inner sep=2.5pt,minimum height=10mm,text width=17mm,fill=white},
  artifact/.style={draw,align=center,inner sep=2.5pt,minimum height=9mm,fill=white},
  flow/.style={-{Latex[length=1.8mm]},semithick},
  card/.style={artifact,align=left,font=\footnotesize,inner sep=3pt,
    execute at begin node={\hyphenpenalty=10000\exhyphenpenalty=10000}},
  note/.style={font=\footnotesize,inner sep=1pt,align=center}]
  % Shading marks the evidence: the agent reply from which the statement was extracted.
  \def\stmtevidence#1#2{\colorbox{black!10}{\parbox[t]{\dimexpr#1-2\fboxsep\relax}{\raggedright#2}}}

  % Discovery traces; the front trace shows an adapted excerpt.
  \node[card,text width=57mm,anchor=north west] (trace) at (0.24,0) {%
    \textbf{Discovery trace (shortened excerpt)}\\[2pt]
    \textbf{User:} My user ID is mia\_li\_3668. Before I request a change, could you explain your rules for changing cabin class on an existing reservation \dots?\\[2pt]
    \stmtevidence{57mm}{\textbf{Agent:} \dots\ Cabin cannot be changed if any flight in the reservation has already been flown.
    In all other cases, reservations, including basic economy, can change cabin without changing the flights.
    Cabin class must remain the same across all flights in the same reservation.
    Changing cabin for just one flight segment is not possible.\ \dots}};
  \draw ([xshift=-1.2mm]trace.north east) |- ([xshift=-1.2mm,yshift=1.2mm]trace.north west)
    |- ([yshift=1.2mm]trace.south west);
  \draw ([xshift=-2.4mm,yshift=1.2mm]trace.north east) |- ([xshift=-2.4mm,yshift=2.4mm]trace.north west)
    |- ([xshift=-1.2mm,yshift=2.4mm]trace.south west);

  % Extraction, one trace at a time.
  \node[operation,text width=15mm,anchor=west] (extract) at ([xshift=5mm]trace.east) {Extract\\statements};
  \node[note,anchor=north] at ([yshift=-1mm]extract.south) {one trace\\at a time};

  % Statement set H; the front card is the statement of the running example.
  \node[card,text width=43mm,anchor=west] (card) at ([xshift=5mm]extract.east) {%
    \textbf{Statement \boldmath$h$}\\[2pt]
    If a cabin change is requested for just one flight segment, the change is not possible.\\[3pt]
    \stmtevidence{43mm}{\textbf{Supporting reply:} \dots\ Changing cabin for just one flight segment is not possible.\ \dots}};
  \draw ([xshift=1.2mm]card.north west) |- ([xshift=1.2mm,yshift=1.2mm]card.north east)
    |- ([yshift=1.2mm]card.south east);
  \draw ([xshift=2.4mm,yshift=1.2mm]card.north west) |- ([xshift=2.4mm,yshift=2.4mm]card.north east)
    |- ([xshift=1.2mm,yshift=2.4mm]card.south east);
  \node[note,anchor=north] at ([yshift=-1mm]card.south) {Statement set $\mathcal H$};

  \draw[flow] (trace.east)--(extract.west);
  \draw[flow] (extract.east)--(card.west);
\end{tikzpicture}
\caption{An LLM extracts statements from the discovery traces, one trace at a time, and stores each statement $h$ with its supporting reply, the agent reply from which it was extracted (shaded).
The trace is a shortened excerpt of a discovery session with the airline agent of $\tau^3$-bench~\cite{tau3bench}, and the statement is one of the statements that the LLM extracted from the agent's reply.}
\label{fig:statement-extraction}
\end{figure}
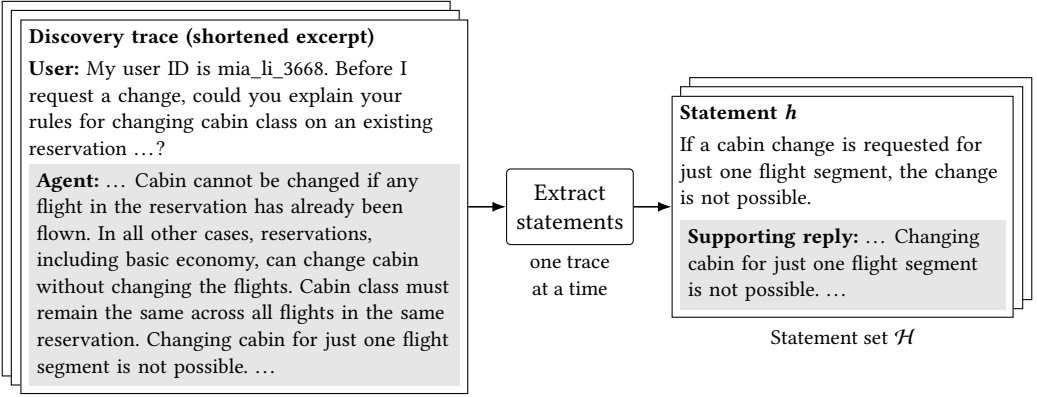

An LLM extracts statements from the agent's replies in the discovery traces and stores each statement together with its supporting reply (Figure~\ref{fig:statement-extraction}, Section~\ref{sec:method-statements}).
A grounding check, also performed by an LLM, removes statements that are not grounded in their supporting replies and orders the rest by the LLM's confidence that an agent must behave as each statement describes.

For a \emph{plain test}, the \emph{generator}, an LLM, receives a statement alone.
It writes at most one plain test for each of a fixed number of highest-ranked statements.
The runner executes these tests, and the judge assigns each a verdict (Section~\ref{sec:prelim-framework}).

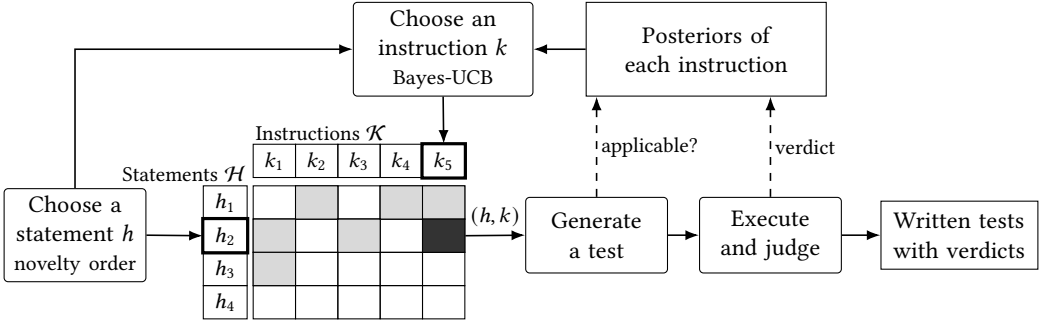
\begin{figure}[t]
\centering
\begin{tikzpicture}[
  x=1cm,y=1cm,font=\small,
  operation/.style={draw,rounded corners=1.5pt,align=center,
    inner sep=2.5pt,minimum height=10mm,text width=17mm,fill=white},
  artifact/.style={draw,align=center,inner sep=2.5pt,minimum height=9mm,fill=white},
  flow/.style={-{Latex[length=1.8mm]},semithick},
  observation/.style={flow,dashed},
  note/.style={font=\footnotesize,inner sep=1pt,align=center},
  obslabel/.style={font=\footnotesize,align=left,inner sep=1.5pt},
  cell/.style={draw,minimum width=5.6mm,minimum height=4.4mm,inner sep=0pt,
    anchor=north west,font=\footnotesize},
  tried/.style={fill=black!15},
  chosen/.style={fill=black!80}]
  % Pair space: rows are statements, columns are instructions.
  \foreach \j in {1,...,5}{
    \node[cell] (k\j) at ({3.30+0.56*(\j-1)},-1.35) {$k_{\j}$};}
  \foreach \i in {1,...,4}{
    \node[cell,minimum width=5.8mm] (h\i) at (2.64,{-1.90-0.44*(\i-1)}) {$h_{\i}$};
    \foreach \j in {1,...,5}{
      \node[cell] (c\i\j) at ({3.30+0.56*(\j-1)},{-1.90-0.44*(\i-1)}) {};}}
  \node[note,anchor=south west] at (k1.north west) {Instructions $\mathcal K$};
  \node[note,anchor=south east] at (h1.north east) {Statements $\mathcal H$};
  % Pairs tried in earlier attempts, and the chosen pair.
  \foreach \c in {c12,c14,c15,c21,c23,c31}{\node[cell,tried] at (\c.north west) {};}
  \node[cell,chosen] at (c25.north west) {};
  \draw[very thick] (h2.south west) rectangle (h2.north east);
  \draw[very thick] (k5.south west) rectangle (k5.north east);

  % Selector: first the statement, then the instruction.
  \node[operation] (choosestatement) at (0.95,-2.56) {Choose a\\statement $h$\\\footnotesize novelty order};
  \node[operation,text width=22mm,minimum height=12.4mm] (chooseinstruction) at (5.82,-0.10)
    {Choose an\\instruction $k$\\\footnotesize Bayes-UCB};
  \draw[flow] (choosestatement.east)--(h2.west);
  \draw[flow] (chooseinstruction.south)--(k5.north);
  \draw[flow] (choosestatement.north) |- (chooseinstruction.west);

  % The generator writes a transformed test for the pair (h,k); the test is executed and judged.
  \node[operation] (generate) at (7.85,-2.56) {Generate\\a test};
  \node[operation] (execute) at (10.15,-2.56) {Execute\\and judge};
  \node[artifact,text width=19mm] (output) at (12.65,-2.56) {Written tests\\with verdicts};
  \draw[flow] (c25.east)--(generate.west) node[midway,above,font=\footnotesize] {$(h,k)$};
  \draw[flow] (generate.east)--(execute.west);
  \draw[flow] (execute.east)--(output.west);

  % The two observations update the posteriors of the chosen instruction.
  \node[artifact,text width=30mm,minimum height=12.4mm] (posteriors) at (9.30,-0.10)
    {Posteriors of\\each instruction};
  \draw[flow] (posteriors.west)--(chooseinstruction.east);
  \draw[observation] (generate.north)--(generate.north |- posteriors.south)
    node[obslabel,pos=0.5,right] {applicable?};
  \draw[observation] (execute.north)--(execute.north |- posteriors.south)
    node[obslabel,pos=0.5,right] {verdict};
\end{tikzpicture}
\caption{One attempt of test transformation.
In each attempt, the selector chooses a statement $h$ by the novelty order and then, by Bayes-UCB, an instruction $k$ whose pair $(h,k)$ is untried (black cell; gray cells are pairs tried in earlier attempts).
After the attempt, the selector updates the posteriors of instruction $k$ with the judge's verdict and with whether the pair is applicable (dashed arrows).}
\label{fig:overview}
\end{figure}

For a \emph{transformed test}, the generator also receives a test transformation instruction from a predefined catalog.
It writes transformed tests in a sequence of \emph{attempts}, which we call \emph{test transformation} (Figure~\ref{fig:overview}).
In each attempt, the \emph{selector} chooses a statement--instruction pair by adaptive pair selection (Sections~\ref{sec:method-instructions} and~\ref{sec:method-selector}). The generator then either writes a transformed test for the pair or reports the pair as not applicable.
The runner executes each transformed test immediately, and the judge assigns it a verdict.
Before the next attempt, the selector updates what it has learned about the chosen instruction using the pair's applicability and the judge's verdict (dashed arrows).
Test generation outputs the plain tests and the remaining transformed tests, each with its verdict.

\subsection{Statements}
\label{sec:method-statements}

A \emph{statement} $h$ is a natural-language hypothesis that describes a single condition and the behavior expected of the agent when that condition holds.
Figure~\ref{fig:statement-extraction} shows a statement extracted from a discovery trace for the running example.
For reservation NO6JO3, the statement describes the boundary side covered by the test in Figure~\ref{fig:test}.

An LLM extracts statements from the discovery traces, one trace at a time.
Because the agent complies with its policy in most conversations (third assumption, Section~\ref{sec:prelim-assumption}), the LLM can treat the behavior stated in the agent's replies as the expected behavior.
From each trace, the LLM writes a statement for each general combination of a condition and an observable behavior---such as a decision or an obligation---that an agent reply states in words.
A reply that states several such combinations therefore yields several statements.
The LLM phrases each statement for a class of entities, such as all reservations, instead of the one entity that a trace discusses.
It leaves out values that do not affect the condition, such as the price of a booking.
Each statement is stored with its \emph{supporting reply}, the agent reply from which it was extracted.
The generator receives each statement and its supporting reply as the target for its plain test and as the starting point for its transformed tests, which can cover other sides or other boundaries (Section~\ref{sec:method-instructions}).
The statements extracted from all traces form the \emph{statement set} $\mathcal H$.

\subsection{Test Transformation Instructions and Statement--Instruction Pairs}
\label{sec:method-instructions}

A \emph{test transformation instruction} $k$, or simply \emph{instruction}, changes one factor of a test, such as the wording of the user's request, a value in that request, or the state of the conversation in which the user makes it.
A \emph{pair} $(h,k)$ combines a statement $h\in\mathcal H$ with an instruction $k\in\mathcal K$, where $\mathcal K$ is the set of instructions.
To write the transformed test for a pair $(h,k)$, the generator receives $h$ together with its supporting reply, the instruction $k$, all discovery traces, and the tests most recently written in the run.
The generator modifies only the step hints, as specified by the instruction; otherwise, it keeps the user request answered by $h$'s supporting reply and other relevant facts unchanged where possible.

A transformed test for $(h,k)$ can cover the side that $h$ describes, the other side of the same boundary, or a boundary that $h$ does not describe.
The statement $h$ specifies what the agent must do only when its condition is met.
If the instruction moves the request to the other side, or to a boundary that $h$ does not describe, the generator's prompt asks it to take the expected behavior from an agent reply in the discovery traces that states the expected behavior for the changed request.
\tool does not verify this expected behavior; our benchmark measures how often the resulting tests are valid (Section~\ref{sec:eval-assessment}).

For the statement in Figure~\ref{fig:statement-extraction}, the instruction to make two related inputs equal can yield a request to change both flights to economy.
This request lies on the other side of the boundary in Figure~\ref{fig:test}, and the agent reply in Figure~\ref{fig:statement-extraction} states the expected behavior.
The supporting reply for the statement ``If the price after the cabin change is lower than the original price, the difference should be refunded'' also states that the cabin class must remain the same across all flights of a reservation.
In one of our runs, the instruction to approve only part of a proposed change yielded a test in which the user proposes to change both business-class flights of a hypothetical reservation to economy, approves the change for the first flight only, and asks whether this partial change is possible.
The test expects the agent to state that this partial change is not possible, so it covers the boundary in Figure~\ref{fig:test}, which the refund statement does not describe.

Each instruction has its own \emph{applicability condition}.
An instruction \emph{applies} to a statement when the statement meets the instruction's applicability condition and the statement's supporting reply---or another agent reply in the discovery traces---justifies the transformation.
For example, the instruction to set a numeric input to zero applies only when the statement concerns a request with a numeric input, such as a refund amount.
The generator decides whether the instruction applies, writes a transformed test for the pair if it does, and otherwise reports the pair as \emph{not applicable}.
The statement in Figure~\ref{fig:statement-extraction} concerns a cabin change, which has no numeric input, so the instruction to set a numeric input to zero does not apply.

\subsection{Pair Selection}
\label{sec:method-selector}

\emph{Pair selection} aims to obtain, in a few attempts, valid boundary tests that cover many boundaries and sides.
However, in our black-box setting, the tester cannot directly observe which tests are valid or which boundaries they cover.
The \emph{selector}, which chooses statement--instruction pairs, can therefore pursue validity and coverage only through proxies.
A transformed test starts from the statement of its pair, and statements with similar text can describe the same boundary.
To cover many boundaries, the selector therefore chooses the statement by the novelty order, which spreads the transformed tests over statements with dissimilar text.
For the transformation instruction, the selector learns from two kinds of feedback that the tester observes after each attempt.
The first is whether the generator finds the pair applicable, which helps the selector obtain tests in fewer attempts.
The second is the judge's verdict on the test, which serves as a proxy for its validity.
In our evaluation, tests with a pass verdict are valid more often than tests with an inconclusive or fail verdict (Section~\ref{sec:discussion}).

Instructions differ widely in how often the generator finds their pairs applicable.
Spreading the attempts evenly over the instructions would waste many attempts on those that the generator rarely finds applicable.
To avoid this waste, the selector learns from earlier attempts which instructions to choose.
Because an instruction alone cannot be used to generate a test and the selector tries each pair at most once, the selector learns about an instruction only from its pairs with other statements.
We assume that the average outcomes of an instruction for earlier statements can guide the selector's choices for later statements.
However, the selector observes outcomes only for the chosen instruction, so within the few attempts of a run it must balance trying rarely chosen instructions against reusing those with good outcomes.
This balance between exploration and exploitation characterizes the \emph{multi-armed bandit problem}~\cite{lai1985asymptotically,lattimore2020bandit}, a sequence of choices among options, the \emph{arms}, whose rewards are unknown in advance.

The selector assigns each instruction a score adapted from the bandit algorithm Bayes-UCB~\cite{kaufmann2012bayesucb}, using upper quantiles of its Beta posteriors (Equation~\ref{eq:instruction-score}).
Algorithm~\ref{alg:transformation} summarizes test transformation with adaptive pair selection, and the following paragraphs describe its parts in detail.

\paragraph{Novelty order.}
\label{sec:method-novelty}
Statements with similar text can describe the same boundary.
For example, the statements ``If a reservation has more than one flight, the cabin class must remain the same across all flights in the reservation'' and ``The cabin class must remain the same across all flights in the same reservation'' both describe the boundary in Figure~\ref{fig:test}.
The \emph{novelty} of a statement $h$ is $\nu(h)=1-\max_{h'\in\mathcal H^{+}}\cos(e_h,e_{h'})$, where $\cos$ is the cosine similarity, $e_h$ is the embedding of the text of $h$, and $\mathcal H^{+}$ is the set of statements that already have a transformed test in the run's output.
If $\mathcal H^{+}$ is empty, $\nu(h)=1$ for every statement, and a statement in $\mathcal H^{+}$ has $\nu(h)=0$.
Under the \emph{novelty order}, the selector chooses the statement with the highest novelty among those it considers in the current block, and it breaks ties at random.

\paragraph{Verdict reward.}
During a run, the tester cannot tell whether a fail verdict reflects a fault in the agent, a defect in the test, or an error by the judge.
The verdict reward is therefore a proxy rather than a measure of validity: it favors instructions whose tests receive a pass verdict, because the agent's behavior matches the test's expected behavior.
This proxy has a cost: it assigns a lower reward to an instruction whose tests expose faults in the agent, and it relies on the third assumption (Section~\ref{sec:prelim-assumption}) to keep this cost small.
The intermediate reward of 0.5 for an inconclusive verdict is a design choice that places this verdict between pass and fail.

\paragraph{Posteriors.}
\label{sec:method-posteriors}
For each instruction $k$, the selector maintains separate Beta posteriors for two quantities, $\theta_k$ and $\mu_k$.
The \emph{applicability probability} $\theta_k$ is the probability that the generator finds a pair with $k$ applicable and writes a transformed test for it.
The quantity $\mu_k$ is the expected verdict reward of a transformed test that the generator writes for a pair with $k$.
When the generator reports a pair with $k$ as not applicable, the selector adds a failure to the posterior for $\theta_k$.
When the generator writes a transformed test for a pair with $k$, the selector adds a success to the posterior for $\theta_k$ and adds $r$ and $1-r$ to the success and failure counts of the posterior for $\mu_k$.

\paragraph{Instruction score.}
At the $i$-th attempt of a run, after choosing the statement $h$ by novelty order, the selector considers each instruction $k$ for which the run has not yet tried the pair $(h,k)$.
Following Bayes-UCB, it computes the quantiles $q_i(\theta_k)$ and $q_i(\mu_k)$ of the posteriors of $\theta_k$ and $\mu_k$ at probability level $1-1/(i+1)$, which increases toward 1 as $i$ grows.
It then chooses the pair whose instruction has the highest \emph{instruction score}:
\begin{equation}
f(k)=q_i(\theta_k)\,q_i(\mu_k)\,\gamma^{m_k},
\label{eq:instruction-score}
\end{equation}
where $m_k$ is the number of earlier attempts on pairs with $k$, and $\gamma\in(0,1)$ is a constant \emph{reuse penalty}.
The product of the two quantiles is a heuristic optimistic estimate of $\theta_k\mu_k$, the expected verdict reward of an attempt on a pair with $k$ when a not-applicable report counts as reward 0; it is not itself a quantile of this product.
The reuse penalty lowers an instruction's score after each earlier attempt on a pair with $k$, spreading attempts across different instructions.
The selector breaks ties in the instruction score at random.

\begin{algorithm}[t]
\caption{Test transformation with adaptive pair selection.}
\label{alg:transformation}
\small
\begin{algorithmic}[1]
\Require statements in their order, instructions $\mathcal K$, reuse penalty $\gamma$, number $n$ of transformed tests
\Ensure transformed tests $T$
\State initialize the posteriors of $\theta_k$ and $\mu_k$ for every $k\in\mathcal K$
\For{$i=1,2,\dots$ until $T$ has $n$ tests or the attempt budget is spent}
  \State $h\gets$ the statement with the highest novelty
  \State $k\gets\arg\max_{k'}\, q_i(\theta_{k'})\,q_i(\mu_{k'})\,\gamma^{m_{k'}}$ over the untried pairs $(h,k')$
  \State $t\gets$ the generator's transformed test for $(h,k)$
  \If{the generator reports $(h,k)$ as not applicable}
    \State add a failure to the posterior of $\theta_k$
  \Else
    \State add a success to the posterior of $\theta_k$; execute and judge $t$
    \State $r\gets$ the verdict reward of $t$; add $t$ to $T$
    \State add $r$ and $1-r$ to the success and failure counts of the posterior of $\mu_k$
  \EndIf
\EndFor
\end{algorithmic}
\end{algorithm}

% ======================================================================
% 4 Evaluation
% ======================================================================
\section{Evaluation}
\label{sec:eval}

We evaluate \tool on four text-based domains from $\tau^3$-bench~\cite{tau3bench}: the airline, retail, and telecom domains from $\tau^2$-bench~\cite{barres2025tau2bench}, and the banking domain from $\tau$-Knowledge~\cite{shi2026tauknowledge}.

\subsection{Research Questions}
\label{sec:eval-rqs}

\paragraph{RQ1 (Effectiveness).}
Across the four domains, how many of \tool's tests are valid boundary tests, how many explicit boundaries and sides do these tests cover, and how many additional explicit boundaries and sides do the transformed tests contribute?

\paragraph{RQ2 (Design choices).}
In the airline and retail domains, how many attempts does adaptive pair selection need to write transformed tests, compared with other selectors, and how do the selectors compare in the validity of these tests and in the additional explicit boundaries and sides that they contribute?

\paragraph{RQ3 (Comparison with baselines).}
Across the four domains, how do \tool's test suites compare with the test suites of two baselines, AgentEval~\cite{agenteval} and a direct generation baseline, in terms of validity and the explicit boundaries and sides covered, and how reliable are the verdicts on the tests of \tool and of the two baselines?

\paragraph{RQ4 (Regression testing).}
In the airline and retail domains, how many seeded policy faults do the test suites of \tool and the two baselines detect as regression tests, and how accurate are verdicts on the faulty agents?

\subsection{Experimental Setup}
\label{sec:eval-setup}

\paragraph{Agents, models, and shared inputs.}
The agent under test is the $\tau^3$-bench agent in each domain, run with DeepSeek V4 Flash 0731.
In each domain, GPT-6 Luna served as the driver for one discovery campaign of 24 sessions, extracted the statements, and performed the grounding check.
The generator is GPT-6 Luna with medium reasoning effort; the judge also uses GPT-6 Luna, and the auditor uses GPT-6.1 Sol.
All runs within a domain share its statements and their order, and all domains share a single catalog of 91 instructions, fixed before any run.

\paragraph{Generated suites.}
A run produces a suite with at most $n$ plain tests and at most $n$ transformed tests, with $n=150$ in RQ1 and $n=50$ in RQ2.
Plain-test generation makes one attempt for each of the first $n$ statements that remain after the grounding check; the retail domain has only 126 such statements.
Test transformation uses one selector and continues until the run has $n$ transformed tests, has made $\min(400,6n)$ attempts, or has passed twice over the blocks; the reuse penalty is $\gamma=0.5$.

\paragraph{Baselines.}
AgentEval's test suite in each domain consists of tests produced by our re-implementation of its boundary generation method~\cite{agenteval}, which uses the same discovery traces and generator model as \tool.
For a fair comparison, we remove AgentEval's limit on the number of tests, allowing it to generate one test for each node and edge of its workflow graph; our runner and judge replace AgentEval's.
We also implement a \emph{direct baseline} to isolate the contribution of statements.
Its generator receives all discovery traces, but no guidance from statements or instructions, and it is instructed to write boundary tests directly from the traces.
It writes up to 150 boundary tests (\emph{direct tests}) and stops early when it determines that the traces support no further tests.

\paragraph{Other selectors.}
\emph{Uniform selection} draws each pair at random from the untried pairs for the current block's statements that are not set aside.
The \emph{static ranking} keeps the novelty order but never updates the posteriors.
Like \tool, the other six selectors learn from the outcomes of earlier attempts.
Four of them keep the novelty order.
\emph{Thompson sampling}~\cite{thompson1933likelihood,russo2018tutorial} replaces the two quantiles in the instruction score with one draw from each posterior.
\emph{UCB1}~\cite{auer2002finite}, \emph{$\varepsilon$-greedy}~\cite{sutton2018reinforcement}, and \emph{Exp3}~\cite{auer2002nonstochastic} choose the instruction using their own rules, without the reuse penalty.
The other two choose the statement and instruction of a pair jointly: \emph{GP-BO} uses Bayesian optimization with a Gaussian process~\cite{shahriari2016bayesian}, and the \emph{LLM selector} asks GPT-6 Luna for an untried pair.

\subsection{Auditor and Measures}
\label{sec:eval-assessment}

\paragraph{Validity labels.}
The auditor reads the agent's artifacts and can search the initial agent state, but it does not see which suite a test belongs to, how the test was generated, or the judge's verdicts.
The auditor's outputs are hidden from the tester.
Using the definitions in Section~\ref{sec:intro}, the auditor labels each test as an explicitly valid boundary test, a valid boundary test that is not explicitly valid, or neither.

\paragraph{Generation measures.}
In RQ1 and RQ3, the \emph{counted tests} are all tests in a suite or generated directly; in RQ2, they are the transformed tests from a run.
Of the $C$ counted tests, $V$ are valid boundary tests and $V_E$ are explicitly valid boundary tests.
The validity rate $V/C$ and the explicit validity rate $V_E/C$ include every counted test, regardless of verdict, because tests with a fail verdict can expose faults.
For each valid boundary test, the auditor identifies every boundary that the test's criteria cover; we call each such boundary, together with the side that the test covers, a \emph{boundary--side mapping}.
A mapping is \emph{explicit} when the auditor supports it with passages from the artifacts.
In a \emph{grouping}, the auditor decides which mappings name the same boundary and the same side.
$U$ is the number of distinct explicit boundaries and $P$ is the number of their distinct sides in the explicit mappings; we report both as medians over three groupings.
$N_n$ counts the attempts of test transformation until a run has $n$ transformed tests, including attempts that end as not applicable.

\paragraph{Execution labels.}
Following AgentEval~\cite{agenteval}, each execution of a valid boundary test receives a label that compares the judge's verdict with the auditor's judgment.
Without seeing the verdict, the auditor judges whether the agent complied with what the artifacts or domain knowledge require for the behavior the test checks.
For this judgment, it reads the execution trace, the agent's tool calls, the changes made to the agent state, and the passages from the agent's artifacts that support the test criteria.
If the agent did not comply, the label is TP for a fail verdict and FN for a pass verdict.
If the agent complied, the label is FP for a fail verdict and TN for a pass verdict.
The label is INC instead of FP or TN when the execution ended before the agent was due to show the behavior that the test checks, or when a criterion demands more than what the artifacts or domain knowledge require.
The label is also INC whenever the judge's verdict is inconclusive.
FP and FN thus expose errors of the judge, and INC exposes problems in the test generator or the runner.
The false alarm rate is $\mathrm{FAR}=\mathrm{FP}/(\mathrm{FP}+\mathrm{TN})$, and the accuracy is $(\mathrm{TP}+\mathrm{TN})$ divided by the number of labeled executions, including those labeled INC.

\paragraph{Executions on the original agent.}
In each domain, the runner executes every test from the suites of \tool and AgentEval once more on the original agent, and the judge assigns a new verdict to each execution.
A human expert (one of the authors) reviewed all 33 of these executions of valid boundary tests with a fail verdict.
For these executions, the label follows the expert's judgment, and the expert takes requirements only from the artifacts, not from domain knowledge.
This review was not blind, because the expert had already seen the verdicts, the auditor's labels, the domain, and the suite for each execution.

\paragraph{Seeded faults and detection.}
For RQ4, we built eight faulty agents in each of the airline and retail domains. Each faulty agent is a copy of the domain agent with one seeded fault: a \emph{policy flip} that rewrites one policy passage to require different behavior.
We defined the faults before running any test on the faulty agents.
For each faulty agent, the runner executes each valid boundary test with an explicit mapping supported by the changed passage. The auditor labels these executions against the original agent's artifacts, without knowing the fault.
A test \emph{detects} a fault when its execution on the original agent has a pass verdict, its execution on the faulty agent has a fail verdict, and the auditor labels the faulty execution TP and, when asked again with a description of the fault, attributes the fail verdict to the fault.
A suite detects a fault when at least one of its tests detects it; its \emph{mutation score} is the proportion of the eight faults in its domain that it detects.
The expert also reviews every execution labeled TP on the faulty agents.

\subsection{Reliability of the Auditor}
\label{sec:eval-reliability}

\begin{table}[t]
\caption{Agreement of each auditor model with the human expert on the airline samples.
In the agreement on the execution labels, the 30 sampled executions with a pass verdict are weighted to represent all 310.
\$/item: cost per labelled test or execution; $\geq$: lower bound, because the charge of one request is unknown.}
\label{tab:reliability}
\footnotesize
\centering
\begin{tabular}{@{}lcccc@{}}
\toprule
 & \multicolumn{2}{c}{Validity labels} & \multicolumn{2}{c}{Execution labels} \\
\cmidrule(lr){2-3}\cmidrule(l){4-5}
Model & Agreement (\%) & \$/item & Agreement (\%) & \$/item \\
\midrule
GPT-6.1 Sol (our auditor) & 85.0 & 0.019 & 99.4 & 0.012 \\
Muse Spark 1.3 & 65.0 & $\geq$0.102 & 99.4 & 0.015 \\
Claude Sonnet 5.5 & 66.7 & 0.031 & 99.4 & 0.019 \\
\bottomrule
\end{tabular}
\end{table}

\begin{table}[t]
\caption{RQ1: test generation metrics of \tool, one run per domain.
In the transformed-test rows, $U$ and $P$ are the extra explicit boundaries and sides that the transformed tests contribute, as the median of the differences over the three groupings, so the rows of a domain need not sum to its All row.
$V/C$ and $V_E/C$ are in percent.
TP: executions of the row's valid boundary tests on the original agent that have a fail verdict and in which the human expert confirms a fault (Section~\ref{sec:eval-assessment}).}
\label{tab:main}
\footnotesize
\setlength{\tabcolsep}{4pt}
\begin{tabular*}{\linewidth}{@{\extracolsep{\fill}}llcccccc@{}}
\toprule
Domain & Test & $C$ & $V/C$ & $V_E/C$ & $U$ & $P$ & TP \\
\midrule
Airline & Plain & 150 & 87.3 & 84.0 & 62 & 70 & 0 \\
 & Transformed & 150 & 89.3 & 82.7 & $+13$ & $+18$ & 1 \\
\cmidrule(l){2-8}
 & All & 300 & 88.3 & 83.3 & 75 & 88 & 1 \\
\midrule
Banking & Plain & 150 & 84.0 & 80.7 & 101 & 103 & 0 \\
 & Transformed & 150 & 85.3 & 82.7 & $+46$ & $+55$ & 1 \\
\cmidrule(l){2-8}
 & All & 300 & 84.7 & 81.7 & 147 & 158 & 1 \\
\midrule
Retail & Plain & 126 & 81.7 & 73.8 & 46 & 54 & 2 \\
 & Transformed & 150 & 86.0 & 80.0 & $+25$ & $+32$ & 7 \\
\cmidrule(l){2-8}
 & All & 276 & 84.1 & 77.2 & 71 & 86 & 9 \\
\midrule
Telecom & Plain & 150 & 71.3 & 60.7 & 58 & 63 & 0 \\
 & Transformed & 150 & 77.3 & 64.7 & $+18$ & $+20$ & 1 \\
\cmidrule(l){2-8}
 & All & 300 & 74.3 & 62.7 & 77 & 84 & 1 \\
\bottomrule
\end{tabular*}
\end{table}

As in AgentEval~\cite{agenteval}, the human expert labels two samples from the airline domain: one test sample and one execution sample.
Table~\ref{tab:reliability} compares these labels with those of our auditor and of two auditors that use the same prompt with models from other families.
Our auditor agrees with the expert on more validity labels than the other two.
The test sample includes 60 tests drawn at random: 20 plain tests and 20 transformed tests from \tool's RQ1 suite, and 20 tests from AgentEval's suite.
The execution sample comes from executions of the valid boundary tests from these suites on the original agent.
It includes all 7 executions with a fail verdict and 30 randomly drawn from those with a pass verdict.

The three models disagree with the expert only on the same 2 of the 37 sampled executions, both with a fail verdict: the models label both TP, and the expert labels one FP and the other INC.

\subsection{Results}
\label{sec:eval-results}

\paragraph{RQ1: Effectiveness.}
Across the four domains, 74.3\% to 88.3\% of \tool's counted tests are valid boundary tests, and 62.7\% to 83.3\% are explicitly valid boundary tests.
The transformed tests are valid boundary tests about as often as the plain tests in the airline and banking domains, and more often in the retail and telecom domains.
In every domain, the transformed tests contribute additional explicit boundaries and sides, and the suites cover 71 to 147 explicit boundaries and 84 to 158 sides (Table~\ref{tab:main}).
On the original agent, the expert confirms a fault of the agent in 12 of the 15 executions of valid boundary tests with a fail verdict.
Although the test suites hold about as many plain tests as transformed tests, the transformed tests account for 10 of these executions and the plain tests for only 2.

\paragraph{RQ2: Design choices.}
The clearest differences between the selectors in Table~\ref{tab:ablation} are in the number of attempts.
These differences come almost entirely from attempts that end as not applicable.
Only 3\% of \tool's attempts end as not applicable in the airline domain and 2\% in the retail domain.

The six other selectors that learn from earlier attempts perform about as well as \tool in validity and in the additional explicit boundaries and sides that their transformed tests contribute, with two exceptions: the LLM selector in the airline domain, whose transformed tests have a lower explicit validity rate and contribute fewer explicit boundaries; and GP-BO in the retail domain, whose transformed tests contribute fewer explicit boundaries and sides.
\tool needs the fewest attempts, or about as few as the best of the other selectors.
Its prior steers it to instructions that apply in most attempts, and its reuse penalty spreads transformed tests over several instructions.
Uniform selection and Exp3 need the most additional attempts in both domains; the LLM selector needs more attempts in the airline domain; and UCB1 and Thompson sampling need more in the retail domain (Table~\ref{tab:ablation}).
The transformed tests of the static ranking have a lower validity rate in the airline domain, and uniform selection contributes fewer explicit boundaries in both domains.

\begin{table}[t]
\caption{RQ2: transformed tests of \tool, of the other selectors, and of the cold start, means $\pm$ standard deviations over three runs.
Except for the Top column, which shows absolute values, the rows below \tool report the difference relative to \tool (the value of the configuration minus that of \tool).
$U$, $P$, $V/C$, and $V_E/C$ as in the transformed-test rows of Table~\ref{tab:main}, with differences in points; $F_V$: valid boundary tests with a fail verdict; Top: percentage of a run's attempts that use the run's most used instruction.}
\label{tab:ablation}
\footnotesize
\centering
\setlength{\tabcolsep}{3.4pt}
\renewcommand{\arraystretch}{0.92}
\sbox0{\begin{tabular}{@{}ll*{7}{c}@{}}
\toprule
Configuration & Changed choice & $N_{50}$ & $V/C$ & $V_E/C$ & $U$ & $P$ & $F_V$ & Top \\
\midrule
\multicolumn{9}{@{}l}{\emph{Airline}} \\
\tool & & $52.0\pm2.6$ & $91.8\pm2.7$ & $84.9\pm4.4$ & $17.8\pm2.2$ & $19.8\pm3.7$ & $0.3\pm0.6$ & $32.7\pm1.6$ \\
\cmidrule{1-9}
Uniform selection & Selector & $+22.7\pm3.1$ & $-1.3\pm1.2$ & $-3.3\pm2.3$ & $-3.3\pm2.3$ & $-3.0\pm4.4$ & $0.0\pm1.0$ & $4.0\pm0.2$ \\
Thompson sampling & Instruction score & $+1.0\pm3.0$ & $-2.7\pm2.3$ & $-4.0\pm2.0$ & $+0.3\pm1.5$ & $+0.7\pm2.5$ & $0.0\pm0.0$ & $21.5\pm6.1$ \\
UCB1 & Instruction score & $0.0\pm2.6$ & $+1.3\pm1.2$ & $-0.7\pm8.3$ & $+1.7\pm3.1$ & $+2.0\pm1.7$ & $-0.3\pm0.6$ & $15.2\pm0.7$ \\
$\varepsilon$-greedy & Instruction score & $+1.0\pm2.0$ & $-2.0\pm3.5$ & $0.0\pm6.9$ & $+0.7\pm4.2$ & $+0.3\pm4.6$ & $0.0\pm0.0$ & $84.4\pm1.6$ \\
Exp3 & Instruction score & $+16.0\pm3.0$ & $+1.3\pm3.1$ & $-2.0\pm11.1$ & $-0.7\pm1.2$ & $+2.0\pm4.4$ & $+1.0\pm1.0$ & $5.8\pm0.3$ \\
Static ranking & Posterior updates & $+2.7\pm2.5$ & $-8.0\pm4.0$ & $-6.7\pm8.1$ & $-1.3\pm2.5$ & $0.0\pm4.4$ & $-0.3\pm0.6$ & $18.9\pm0.7$ \\
GP-BO & Selector & $+0.3\pm2.1$ & $-0.7\pm4.2$ & $+0.7\pm4.2$ & $-0.7\pm3.2$ & $0.0\pm4.6$ & $+0.3\pm1.5$ & $94.9\pm1.2$ \\
LLM selector & Selector & $+6.3\pm0.6$ & $-7.3\pm8.3$ & $-14.0\pm10.0$ & $-2.3\pm1.2$ & $-2.0\pm4.0$ & $-0.3\pm0.6$ & $11.4\pm0.9$ \\
\midrule
\multicolumn{9}{@{}l}{\emph{Retail}} \\
\tool & & $52.8\pm2.0$ & $88.3\pm3.3$ & $79.2\pm1.3$ & $16.6\pm0.7$ & $21.8\pm0.9$ & $0.5\pm0.5$ & $32.2\pm1.2$ \\
\cmidrule{1-9}
Uniform selection & Selector & $+17.7\pm6.8$ & $-0.7\pm1.2$ & $0.0\pm4.0$ & $-2.7\pm1.2$ & $-5.7\pm1.2$ & $0.0\pm1.7$ & $4.3\pm0.3$ \\
Thompson sampling & Instruction score & $+5.7\pm2.3$ & $+1.3\pm4.2$ & $+1.3\pm4.2$ & $-0.7\pm6.8$ & $-2.7\pm7.6$ & $-0.7\pm1.2$ & $15.6\pm5.5$ \\
UCB1 & Instruction score & $+8.3\pm2.1$ & $+0.7\pm3.8$ & $-3.7\pm1.5$ & $-0.3\pm0.8$ & $-2.5\pm1.0$ & $-0.7\pm0.6$ & $13.1\pm0.8$ \\
$\varepsilon$-greedy & Instruction score & $+1.7\pm0.6$ & $-4.0\pm10.6$ & $-6.0\pm13.1$ & $-1.0\pm4.6$ & $-0.7\pm4.7$ & $0.0\pm1.0$ & $85.0\pm3.2$ \\
Exp3 & Instruction score & $+15.3\pm4.0$ & $+1.3\pm6.4$ & $+3.3\pm9.0$ & $-1.0\pm1.7$ & $+1.3\pm1.2$ & $+0.7\pm1.2$ & $5.5\pm0.9$ \\
Static ranking & Posterior updates & $-1.0\pm4.0$ & $-2.7\pm9.0$ & $+1.3\pm7.0$ & $+0.3\pm3.1$ & $+1.0\pm1.0$ & $+0.7\pm1.2$ & $12.8\pm0.7$ \\
GP-BO & Selector & $+1.0\pm2.6$ & $-1.3\pm3.1$ & $+1.3\pm1.2$ & $-3.0\pm2.0$ & $-3.0\pm0.0$ & $-0.7\pm0.6$ & $83.7\pm15.3$ \\
LLM selector & Selector & $+2.3\pm2.5$ & $+1.3\pm2.3$ & $+1.3\pm1.2$ & $-1.0\pm2.0$ & $-3.0\pm3.6$ & $+1.0\pm1.7$ & $16.4\pm5.8$ \\
\bottomrule
\end{tabular}}%
\ifdim\wd0>\linewidth\resizebox{\linewidth}{!}{\usebox0}\else\usebox0\fi
\end{table}

\begin{table}[t]
\caption{RQ3: the test suites of \tool from RQ1 and AgentEval's suite in each domain.
Execution: the executions of the valid boundary tests on the original agent; TP as in Table~\ref{tab:main}; INC: proportion of the labelled executions that have the label INC.
$V/C$, $V_E/C$, INC, Acc.\ (accuracy), and FAR (false alarm rate) are in percent.
The auditor groups the mappings of the two methods separately.
Without a label: one execution of \tool's tests with an incomplete record in each of the banking and telecom domains, and 3 executions of \tool's telecom tests with a pass verdict for which the auditor cannot determine whether the agent complied.}
\label{tab:agenteval}
\footnotesize
\centering
\setlength{\tabcolsep}{3pt}
\begin{tabular}{@{}ll*{9}{c}@{}}
\toprule
 & & \multicolumn{5}{c}{Generation} & \multicolumn{4}{c}{Execution} \\
\cmidrule(lr){3-7}\cmidrule(l){8-11}
Domain & Method & $C$ & $V/C$ & $V_E/C$ & $U$ & $P$ & TP & INC & Acc. & FAR \\
\midrule
Airline & \tool & 300 & 88.3 & 83.3 & 75 & 88 & 1 & 8.7 & 91.3 & 0.0 \\
 & AgentEval & 90 & 82.2 & 55.6 & 71 & 78 & 3 & 6.8 & 87.8 & 1.6 \\
 & Direct baseline & 126 & 72.2 & 59.5 & 67 & 74 & 0 & 11.0 & 86.8 & 2.5 \\
\midrule
Banking & \tool & 300 & 84.7 & 81.7 & 147 & 158 & 1 & 17.8 & 81.0 & 0.5 \\
 & AgentEval & 91 & 90.1 & 68.1 & 82 & 101 & 1 & 8.5 & 84.1 & 4.2 \\
 & Direct baseline & 104 & 84.6 & 78.8 & 84 & 89 & 1 & 4.5 & 94.3 & 0.0 \\
\midrule
Retail & \tool & 276 & 84.1 & 77.2 & 71 & 86 & 9 & 20.7 & 79.3 & 0.0 \\
 & AgentEval & 83 & 90.4 & 66.3 & 70 & 77 & 1 & 4.0 & 92.0 & 2.9 \\
 & Direct baseline & 83 & 84.3 & 71.1 & 60 & 62 & 4 & 11.4 & 85.7 & 3.4 \\
\midrule
Telecom & \tool & 300 & 74.3 & 62.7 & 77 & 84 & 1 & 21.9 & 78.1 & 0.0 \\
 & AgentEval & 86 & 83.7 & 47.7 & 64 & 65 & 1 & 15.3 & 79.2 & 3.4 \\
 & Direct baseline & 96 & 77.1 & 52.1 & 49 & 51 & 3 & 18.9 & 77.0 & 0.0 \\
\bottomrule
\end{tabular}
\end{table}

\paragraph{RQ3: Comparison with baselines.}
In AgentEval's test suites (Table~\ref{tab:agenteval}), 22.0\% to 36.0\% of the counted tests are valid boundary tests that are not explicitly valid, compared with 3.0\% to 11.7\% in \tool's test suites.
On the original agent, four of AgentEval's eight executions labeled FP---including the only one in the airline domain---are executions of such tests.

The auditor labels the direct tests together with the plain tests: 52.1\% to 78.8\% of the direct tests are explicitly valid boundary tests, compared with 60.0\% to 83.3\% of the plain tests.
For equal numbers of tests, however, the direct tests cover as many or more explicit boundaries (49 to 84), compared with 38 to 73 for the plain tests.

In all four domains, \tool's test suites have a higher explicit validity rate (Table~\ref{tab:agenteval}); AgentEval's test suites have a higher validity rate in the banking, retail, and telecom domains.
In most of AgentEval's valid boundary tests that are not explicitly valid, a criterion describes a general conversational behavior that the artifacts do not define.
With more than three times as many tests, \tool's test suites cover more sides in every domain, with more explicit boundaries in the banking and telecom domains and about as many in the airline and retail domains.
On the original agent, \tool's test suites have the lower false alarm rate in every domain, although in the airline domain the difference comes from a single execution.
The expert confirms an agent fault in all 12 executions of \tool's tests that the auditor labels TP, compared with 6 of the 13 TP-labeled executions of AgentEval's tests.
A larger proportion of executions of \tool's tests is labeled INC, mostly because the judge returns inconclusive verdicts on plain tests.
This is a weakness of \tool's test suites; as a result, they have lower accuracy in the banking, retail, and telecom domains.
Compared with the direct tests, the plain tests are explicitly valid more often, but for equal numbers of tests, the direct tests cover as many or more explicit boundaries.
However, the direct baseline stops early, after 126, 104, 83, and 96 tests in the airline, banking, retail, and telecom domains, respectively, when it finds that the traces support no further tests.

\begin{table}[t]
\caption{RQ4: the airline and retail test suites of Table~\ref{tab:agenteval} on the eight faulty agents of each domain.
TP to INC: the auditor's labels of the executions of the valid boundary tests, summed over the faulty agents; Conf.: executions labelled TP in which the expert confirms a fault; Det.: detected faults (of 8); MS: mutation score in percent.}
\label{tab:mutants}

\footnotesize
\centering
\sbox0{\begin{tabular}{@{}ll*{10}{c}@{}}
\toprule
Domain & Method & TP & FP & TN & FN & INC & FAR & Acc. & Conf. & Det. & MS \\
\midrule
Airline & \tool & 38 & 0 & 23 & 3 & 25 & 0.0 & 68.5 & 37 & 8 & 100.0 \\
 & AgentEval & 14 & 0 & 39 & 2 & 5 & 0.0 & 88.3 & 12 & 5 & 62.5 \\
 & Direct baseline & 10 & 0 & 17 & 0 & 7 & 0.0 & 79.4 & 10 & 6 & 75.0 \\
\midrule
Retail & \tool & 17 & 0 & 42 & 1 & 10 & 0.0 & 84.3 & 16 & 4 & 50.0 \\
 & AgentEval & 10 & 0 & 34 & 5 & 4 & 0.0 & 83.0 & 8 & 2 & 25.0 \\
 & Direct baseline & 6 & 0 & 27 & 1 & 4 & 0.0 & 86.8 & 4 & 1 & 12.5 \\
\bottomrule
\end{tabular}}%
\ifdim\wd0>\linewidth\resizebox{\linewidth}{!}{\usebox0}\else\usebox0\fi
\end{table}

\paragraph{RQ4: Regression testing.}
In the airline domain, \tool's test suite detects all eight seeded faults, the direct baseline's test suite detects six, and AgentEval's test suite detects five (Table~\ref{tab:mutants}).
AgentEval's test suite misses three airline faults mainly because few of its tests cover a boundary affected by these faults: 0, 3, and 1 of its tests do, compared with 7, 13, and 13 of \tool's tests.

In the retail domain, \tool's test suite detects four of the eight seeded faults, AgentEval's test suite detects two of these four, and the direct baseline's test suite detects one.
All three test suites miss the other four faults, whose policy flips concern authenticating a user by a stated user ID, transferring to a human agent, the accepted reasons for canceling an order, and exchanging a delivered item only for a different option of the same product.

In the airline domain, more executions of \tool's tests are labeled INC, all because the judge returns inconclusive verdicts, so its suite has lower accuracy.
In the retail domain, the three test suites have about the same accuracy (Table~\ref{tab:mutants}).

% ======================================================================
% 5 Discussion, 6 Related Work, 7 Conclusion
% ======================================================================
\section{Discussion and Threats to Validity}
\label{sec:discussion}
\label{sec:threats}

\paragraph{Both sides of a boundary.}
Because a boundary has two sides, $P-U$ in one grouping is the number of explicit boundaries for which the tests cover both sides.
In the airline, banking, retail, and telecom domains, $P$ exceeds $U$ by 13, 11, 15, and 7 for \tool's test suites (Table~\ref{tab:main}) and by 7, 19, 7, and 1 for AgentEval's test suites (Table~\ref{tab:agenteval}).
Thus, in each domain, both methods cover less than a quarter of their explicit boundaries on both sides.
Tests that cover only one side of a boundary cannot show whether the agent distinguishes the two sides. Future work should address that limitation.

\paragraph{Internal validity.}
Because each run requires thousands of LLM calls, experiment costs limited us to three runs per configuration in RQ2 and one test suite per method and domain in RQ1, RQ3, and RQ4, which is too few for significance tests.
In RQ2, however, the three runs of \tool vary little, with standard deviations of at most 2.6 attempts in $N_{50}$ and 4.4 points in validity rates (Table~\ref{tab:ablation}), suggesting that a single run of \tool is representative.
A test that exposes an agent fault receives a reward of 0, so the selector may choose that instruction less often.
In our experiments, however, transformed tests account for 10 of the 12 executions on the original agent in which the expert confirms a fault.
Our re-implementation of AgentEval may differ from the original, and our runner and judge may change the verdicts on its tests.
Even without a test limit, AgentEval generates at most one test per node and edge of its workflow graph, so its smaller test suites are likely to cover fewer explicit boundaries and sides and detect fewer faults.

\paragraph{Construct validity.}
Validity labels, boundary and side counts, and most execution labels come from the auditor, and LLMs used as judges can show biases~\cite{zheng2023judge}.
Absolute validity rates depend on the auditor model, but the test suites compared within a domain share the same auditor.
The auditor alone decides all other labels and attributes fail verdicts to seeded faults.

\paragraph{External validity.}
We use four diverse domains of $\tau^3$-bench, one agent model, one generator model, one discovery campaign per domain, and one catalog of 91 instructions.
To limit experiment costs, RQ2 and RQ4 use only the airline and retail domains.
The seeded faults rewrite only policy text in the agent's prompt.

\section{Related Work}
\label{sec:related}

\paragraph{Testing LLM-based agents.}
IntellAgent generates tests from the policies in an agent's system prompt~\cite{levi2025intellagent}, the Agent-Testing Agent analyzes the agent's code~\cite{komoravolu2026ata}, and SpecOps sets up the agent's environment before each test~\cite{ahmed2026specops}.
Sensei checks a chatbot's conversations with an LLM user simulator against assertions written by human testers~\cite{delara2026sensei}.
Tracer explores a chatbot as a black box and generates profiles of the simulated user's goals and interaction style~\cite{sotillo2026tracer}.
AgentEval generates tests for the nodes and edges of a workflow graph mined from discovery traces~\cite{agenteval}.
As in AgentEval, \tool's tester observes only user messages and agent replies and starts every session from the same initial agent state.
\tool replaces AgentEval's nodes and edges with extracted statements, so the generator receives the expected behavior directly rather than inferring it.

\paragraph{Expected behaviors.}
In CheckList~\cite{ribeiro2020checklist}, AdaTest~\cite{ribeiro2022adatest}, and Weaver~\cite{yang2023weaver}, human testers write, filter, or label expected test behaviors.
Specification mining infers likely invariants~\cite{ernst2001invariants} or state machines~\cite{ammons2002mining,lorenzoli2008gktail} from a program's observed executions.
AgentPex checks an agent's traces against specifications extracted from the agent's prompt~\cite{sharma2026agentpex}.
Because the agent complies with its policy in most conversations (third assumption, Section~\ref{sec:prelim-assumption}), \tool extracts statements from the agent's replies in the discovery traces and generates new tests from them.
A statement is closer to a likely invariant than to a state machine because it describes a single condition and the behavior expected when that condition holds.

\paragraph{Boundary testing.}
Domain testing~\cite{white1980domain,clarke1982domain} and boundary value analysis~\cite{hierons2006bva} place tests near the boundaries between groups of inputs defined by a program's code or specification.
AutoBVA~\cite{dobslaw2023autobva} and SETBVE~\cite{akbarova2026setbve} detect black-box program boundaries from small distances between inputs and large distances between outputs.
For the boundaries that \tool tests, two identical user requests (distance zero) can lie on different sides, because the side can also depend on what the conversation has established (Section~\ref{sec:prelim-boundary}).

\paragraph{Test purposes, patterns, and partitions.}
A statement resembles a test purpose, which specifies a set of tests in abstract terms~\cite{ledru2001testpurposes}.
The applicability condition of an instruction corresponds to the context of a test-selection pattern, which suggests tests for the risks in that context~\cite{howden2005testselection}.
Category-partition testing combines choices for categories of parameters and environment conditions, under constraints that human testers list in advance~\cite{ostrand1988categorypartition}.
A pair combines one choice from each of two categories: statements and instructions.
In place of such constraints, the generator decides whether the instruction applies to the statement and otherwise reports the pair as not applicable.

\paragraph{Metamorphic testing.}
Metamorphic testing checks relations among the inputs and outputs of several executions~\cite{chen2018metamorphic}, for example, that a chatbot behaves the same way after keywords in a request are replaced with synonyms~\cite{bozic2019chatbotmt}.
Several instructions in \tool also change a request's wording, but the judge checks each transformed test against its own criteria in a single execution.

\paragraph{Adaptive selection.}
Search-based testing guides test generation by a fitness function~\cite{mcminn2004searchbased}, for example toward coverage targets in EvoSuite~\cite{fraser2011evosuite} and DynaMOSA~\cite{panichella2018dynamosa} or toward inappropriate responses of LLM applications in STELLAR~\cite{sorokin2026stellar}.
In our black-box setting, \tool's tester cannot observe which tests are valid or which explicit boundaries they cover.
The selector therefore learns from whether the generator writes a test and from the judge's verdict, using a bandit whose arms are the instructions.
Greybox fuzzers divide effort among seeds with power schedules~\cite{bohme2016aflfast,bohme2020entropic} or bandits~\cite{yue2020ecofuzz,luo2024tscheduler}, and other fuzzers learn which mutation operators to apply~\cite{lyu2019mopt,karamcheti2018thompson}.
COLEMAN prioritizes tests in continuous integration with a bandit that learns from test failures~\cite{pradolima2022coleman}.
Unlike COLEMAN, \tool's verdict reward favors instructions whose tests receive a pass verdict, because a fail verdict can also result from a defective test or a judge error (Section~\ref{sec:method-selector}).

\paragraph{Adaptive random testing.}
Adaptive random testing (ART) aims to spread test inputs evenly over the input domain~\cite{chen2004art} and has recently been adapted to LLM-based applications~\cite{yoon2025adaptive}.
Analogously, \tool's novelty order disperses transformed tests across statements that are textually dissimilar Similarly, \tool's novelty order disperses transformed tests across textually dissimilar statements (Section~\ref{sec:method-novelty}).

\section{Conclusion}
\label{sec:conclusion}

We presented \tool, which extracts statements from an agent's replies in discovery traces and uses them to guide boundary-test generation.
In addition to plain tests for individual statements, \tool generates transformed tests from statement--instruction pairs.
Because the number of possible pairs far exceeds the attempts available for a run, an adaptive selector prioritizes pairs that yield valid tests.
Across the four $\tau^3$-bench domains, \tool's test suites contain a higher proportion of explicitly valid boundary tests than those of AgentEval and a direct baseline.
In every domain, transformed tests cover explicit boundaries that plain tests miss, and they account for most executions on the original agent in which an expert confirms a fault.
As regression tests, \tool's test suites detect all seeded policy faults in the airline domain and half in the retail domain, outperforming both baselines.
Future work includes generating tests for both sides of more explicit boundaries and reducing the judge's inconclusive verdicts, which currently lower test-suite accuracy.

\section*{Data Availability}
The code, prompts, and data used in this paper will be made publicly available upon acceptance.

\begin{acks}
This work has emanated from research jointly funded by Taighde \'{E}ireann--Research Ireland under Grant Number 13/RC/2094\_2, and by Genesys Cloud Services, Inc.
\end{acks}

\bibliographystyle{ACM-Reference-Format}
\bibliography{references}

\end{document}